\documentclass[12pt]{article}

\usepackage{graphicx}
\usepackage{amsmath,amssymb}

\makeatletter \@addtoreset{equation}{section}

\renewcommand{\thefootnote}{\fnsymbol{footnote}}
\newcommand{\be}{\begin{equation}}
\newcommand{\ee}{\end{equation}}
\newcommand{\bear}{\begin{eqnarray}}
\newcommand{\eear}{\end{eqnarray}}
\newcommand{\ba}{\begin{array}}
\newcommand{\ea}{\end{array}}

 \newcommand{\ra}{{\rightarrow}}
 \newcommand{\del}{{\delta}}

\newcommand{\nn}{\nonumber \\}

\newcommand{\tr}{{\rm tr}}
\newcommand{\Tr}{{\rm Tr}}

\newcommand{\bsubeq}{\begin{subequations}}
\newcommand{\esubeq}{\end{subequations}}

\newcommand{\g}{{\rm g}}

\def\tr{{\rm tr}}\def\Tr{{\rm Tr}}
\def\del{\partial}
\def\calO{\cal O}

\def\pr{Phys. Rev.}\def\prl{Phys. Rev. Lett.}

\def\la{\langle}\def\ra{\rangle}

\def\bi{\bibitem}

\def\lsim{\mathrel{\rlap{\lower3pt\hbox{\hskip1pt$\sim$}}
     \raise1pt\hbox{$<$}}} 
\def\gsim{\mathrel{\rlap{\lower3pt\hbox{\hskip1pt$\sim$}}
     \raise1pt\hbox{$>$}}}
\def\HLS1{HLS$_1$}

\DeclareFontFamily{U}{rsf}{}
\DeclareFontShape{U}{rsf}{m}{n}{
  <5> <6> rsfs5 <7> <8> <9> rsfs7 <10-> rsfs10}{}
\DeclareMathAlphabet\Scr{U}{rsf}{m}{n}

\begin{document}
\begin{titlepage}
\vfill
\begin{flushright}
{\tt\normalsize KIAS-P07017}\\
{\tt\normalsize PNUTP-07/A03}\\
{\tt\normalsize hep-th/yymmnnn}\\
\end{flushright}
\vfill
\begin{center}
{\Large\bf Dynamics of Baryons from String Theory\\ and Vector Dominance }

\vfill

Deog Ki Hong$^{\diamondsuit}$\footnote{\tt dkhong@pusan.ac.kr} ,
Mannque Rho$^{\heartsuit}$\footnote{\tt Mannque.Rho@cea.fr} ,
Ho-Ung Yee$^{\spadesuit}$\footnote{\tt ho-ung.yee@kias.re.kr} ,
and Piljin Yi$^{\spadesuit}$\footnote{\tt piljin@kias.re.kr}

\vskip 5mm
$^{\diamondsuit}${\it Department of Physics, Pusan National University, Busan 609-735, Korea }
\vskip 1mm
$^{\heartsuit}${\it  Service de Physique Th\'eorique, CEA Saclay, 91191
Gif-sur-Yvette, France}\\
\vskip 1mm
$^{\spadesuit}${\it  School of Physics, Korea Institute for Advanced Study, Seoul 130-722, Korea}

\end{center}
\vfill

\begin{abstract}
\noindent We consider a holographic model of QCD from string theory,
{\it \`a la } Sakai and Sugimoto, and study baryons. In this model,
mesons are collectively realized as a five-dimensional $U(N_F)=U(1)\times SU(N_F)$
Yang-Mills field and baryons are classically identified as
$SU(N_F)$ solitons with a unit Pontryagin number and $N_c$ electric
charges. The soliton is shown to be very small in the large 't Hooft coupling limit,
allowing us to introduce an effective field ${\cal B}$. Its coupling
to the mesons are  dictated by the soliton structure, and
consists of a direct magnetic coupling to the $SU(N_F)$ field strength
as well as a  minimal coupling to the $U(N_F)$ gauge field. Upon the dimensional reduction,
this effective action reproduces all interaction terms between nucleons and
an infinite tower of mesons in a manner consistent with the large $N_c$
expansion. We further find  that all electromagnetic interactions, as
inferred from the same effective action via a holographic prescription,
are mediated  by an infinite tower of vector mesons,
rendering the baryon electromagnetic form factors $completely$
vector-dominated as well. We estimate nucleon-meson couplings
and also the anomalous magnetic moments, which compare well with nature.
\end{abstract}

\vfill
\end{titlepage}

\tableofcontents\newpage
\renewcommand{\thefootnote}{\#\arabic{footnote}}
\setcounter{footnote}{0}

\section{Introduction}

The recent development in applying the concept and the methodology
of AdS/CFT duality \cite{Maldacena:1997re} to low-energy hadron
dynamics, referred to as  the holographic QCD or AdS/QCD, brings out
two related issues from opposite directions, one top-down from string
theory \cite{sakai-sugimoto,Karch:2002sh} and the other bottom-up
from low-energy chiral effective field theory of mesons and baryons
\cite{son-stephanov,Erlich:2005qh,Hong:2006ta}.

{}From the string theory point of view, what one is interested in is
to assess to what extent the gravity theory in the bulk sector in a
controlled weak coupling limit can address, via duality, the
strongly coupled dynamics of QCD and if so, how well and how far. In
this respect, the aim there is to ``post-dict" what is established
in low-energy hadron dynamics, and try to reproduce what has been
well understood in low-energy effective theories. The principal goal
here is to establish its raison-d'\^etre in the strong interaction
sector. On the other hand, from the low-energy effective theory
perspective on which we will elaborate in some detail below, the aim
is, if it is firmly established that the holographic QCD has definite connection
to real QCD, whether it can make clear-cut and falsifiable
predictions on processes which are difficult to access by QCD proper.

A notable example of this sort is the prediction by AdS approaches
of low viscosity-entropy ratio \cite{Kovtun:2004de} and also of
low elliptic flow in matter at high temperature above the
chiral restoration point~\cite{Janik:2005zt}, which
is presumed to be observed at RHIC. Given the complete inability
of the QCD proper to handle this regime, this development gives a
hope that the holographic approach could provide a powerful tool going
beyond perturbative QCD and elucidate strongly interacting matter
under extreme conditions that are otherwise inaccessible, such as
the phenomenon of jet-quenching \cite{Sin}. Another outstanding immediate
challenge to AdS approaches is to identify and elucidate the
degrees of freedom figuring just below (in the Nambu-Goldstone
phase) and just above (in the Wigner-Weyl phase) $T_c$, the chiral
transition temperature presumed to have been probed at
RHIC~\cite{BGR06}. At present, however, in the paucity of better
understanding, it is not clear whether the current ``explanation"
of the properties of quark-gluon plasma at RHIC reflects directly
certain specific properties of nonperturbative QCD or whether they
are simply in a same universality class unspecific to dynamics.
For instance, recent works suggest that the prediction of viscosity-entropy
ratio could be common to all AdS-based models regardless of
details~\cite{bucheletal}.

Another example of this sort is the exploitation of the conformal
structure of AdS/CFT to deduce the analytic form of the
frame-independent light-front wavefunctions of hadrons which could
allow the computation of various observables that are found to be
difficult to obtain in QCD itself~\cite{brotskyetal}.

In this paper, we would like to zero in on a more specific set of
problems that are typically of strong-coupling QCD and are very
difficult to access by established QCD techniques, namely chiral
dynamics of hadrons, in particular baryons, at low energy.
Unlike pions, which are relatively
well-understood from the chiral Lagrangian approach to QCD,
baryons remain more difficult to pin down. This may account for
the reason why in the chiral lagrangian approach, baryons are
either put in by hand as  point-like objects or built up as
solitons (i.e., Skyrmions) from mesons. The former suffers from
the lack of theoretical justification as a local field when the
energy scale reaches the inverse of its Compton wavelength as
evidenced in the growing number of unknown parameters, while the
latter in its simplest approximation does not fare well in
phenomenology. Attempts to marry the two pictures are often
difficult, given the relatively large size of the Skyrmion.

This work was motivated by an astute modelling of chiral dynamics
within the framework of AdS/CFT by Sakai and
Sugimoto~\cite{sakai-sugimoto} that correctly describes the
spontaneous breaking of chiral $U(N_F)_L\times U(N_F)_R$ to the
diagonal subgroup $U(N_F)_{L+R}$. For our purpose, the most salient
feature of the holographic model of Sakai and Sugimoto (SS for
short) is that the entire tower of vector mesons plus the pions are
built into a single $U(N_F)$ gauge field in five-dimensions,
immensely simplifying possible interaction structures among mesons,
and eventually with baryons as well. This also implies that the
low-energy chiral dynamics incorporating the ``hidden local gauge
symmetry" (HLS) is manifest in five dimensions. The $U(N_F)$ gauge
field is supported by $N_F$ D8 branes compactified on $S^4$ while
the strongly coupled $SU(N_c)$ dynamics is hidden in the background
AdS-like geometry.

The effective chiral theory, defined at a KK scale, $M_{KK}$ (commensurate
with the chiral scale $\Lambda_\chi\sim 4\pi f_\pi$) is valid and
justified in the limit of large 't Hooft coupling constant
$\lambda=g_{YM}^2N_c$ and large $N_c$. Surprisingly even in this
limit, the SS model has been shown to possess the power to reproduce
rather well  most of the low-energy hadron properties in the meson
sector that are highly non-perturbative, such as, for example,
soft-pion theorems, KSRF relations, various sum rules etc. Most
notable among what has been obtained is that $all$ hadron processes
involving mesons, both normal (e.g., $\pi$-$\pi$ scattering) and
anomalous (e.g., $\pi^0\rightarrow 2\gamma$, $\omega\rightarrow
3\pi$ etc), are {\it vector-dominated} with all the members of the
infinite tower participating non-trivially in the process, including
the well-established vector dominance of the pion EM form factor.

Now given an effective theory that captures the physics of the
meson sector, one immediate question is how the baryons figure in
the story and how well the picture approximates static and dynamic
properties of baryons. A related question is whether or not the
vector dominance which holds $naturally$ in the meson sector also
holds with baryons. This question has a bearing on the concept of
``universality" that has played an important role in the history
of vector dominance in hadron physics. As described in detail in
what follows, a baryon in the SS model is a soliton with
instanton-like configuration in a five-dimensional Yang-Mills
action, which encodes the winding number of the four-dimensional
Skyrmion made up not only of the pion field but also of an
infinite tower of vector mesons.

In fact, perhaps the most appealing possibility for the holographic QCD to
unravel something truly novel in low-energy hadron dynamics is in
the baryon structure -- which is the principal subject of this
article. In the past, several authors~\cite{vector-skyrmion}
studied Skyrmions with the HLS Lagrangian containing the lowest
vector mesons, $\rho$ and $\omega$, of
\cite{bandoetal}~\footnote{Some works include the $a_1$ meson as
well, but the idea is essentially the same as without it.} as
merely an {\it alternative or improved} description of the same
soliton given by the Skyrme model with the pion
hedgehog~\cite{skyrme,ANW}. The essential idea was that vector
mesons, in particular the $\rho$ meson, could replace the Skyrme
quartic term in the role of stabilizing the soliton~\footnote{We
will argue in Section 8 that this idea is not correct.}. It was
only recently suggested that hidden local fields bring a
drastically different or novel aspect to the soliton structure of
baryons~\cite{brihayeetal,rho,Nawa:2007gh}. Indeed what we have
found is that the instanton baryon, which is a Skyrmion with an
infinite tower of hidden local fields, presents an aspect of
baryons hitherto left largely unexplored.

A major part of this paper will be devoted to understanding the
simplest of static properties, and subsequently the chiral dynamics
of the baryons realized as five-dimensional solitons. One
consequence of the fully five-dimensional picture is that, in the
large 't Hooft coupling, the instanton size is so small to be
amenable to a simple effective field theory approach. Our strategy
in uncovering the dynamics of baryons relies on an effective field
theory of the small instanton in the five dimensional setting. The
quantum numbers of the small instanton are commensurate with those
of the Skyrmion, except that it naturally and minimally couples to
five-dimensional $U(N_F)$ gauge fields instead of to
four-dimensional $SU(N_F)$ pion fields. This results in a simple
five-dimensional Dirac field representing baryons, minimally coupled
to the $U(N_F)$ gauge field. While the instanton size is small, on
the other hand, the long distance power-like tail cannot be ignored
and leads to a higher-dimensional coupling between the $SU(N_F)$
field strength and the Dirac field whose coupling strength is
determined by the size of the instanton. It is plausible that this
picture can justify the long-standing tradition -- recently given a
support in terms of chiral perturbation theory -- in nuclear physics
where the nucleon is considered as a point-like object and its
finite size effects are taken into account via ``meson cloud."

At the end of the day, what will have transpired is that these two
simple and explicitly computable five-dimensional interaction terms
in the baryon effective action encode all the four-dimensional
meson-baryon interactions, up to quadratic order in the baryon
field. This includes the pions, the entire tower of vector mesons
and axial vector mesons, once and for all, and also incorporates
iso-scalar and iso-vector mesons on equal
footing. Needless to say, this will result in a large number of
predictions on various meson-baryon couplings, and more indirectly,
various electromagnetic interactions.

While the photon field is not present among the degrees of freedom
in this model, the electromagnetic current can also be extracted
following the general prescription of AdS/CFT. An interesting
outcome of this investigation is that, although the effective action
approach in five dimensions predicts a minimal coupling between
photon field and the baryon, a mixing between massive vector mesons
and the photon field effectively replaces this with an infinite
number of vector mesons coupling to the baryons. The resulting
electromagnetic form factors show a complete vector dominance in the
sense that all electromagnetic interactions are mediated by exchange
of vector mesons, generalizing old notion of vector dominance by the
lightest vector mesons. In particular, full vector dominance is
recovered in the electromagnetic form factors of the nucleon in the
same fashion as in the pion.

We will also discuss subleading $1/N_c$
corrections and compare these findings against experimental values.
Throughout the derivation of the effective action, we stay in the
regime of large $N_c$ and large `t Hooft coupling $g^2_{YM}N_c$
where the size of baryon is small enough to justify this approach.
On the other hand, the realistic regime of $N_c=3$ QCD with the pion
decay constant $f_\pi\sim$ 93 MeV demands $g^2_{YM}N_c\sim 17$, which
is not large enough. The baryon size is difficult to estimate in
this regime but is clearly of the same order as $1/M_{KK}$. To avoid
the difficulties associated with the latter, we take the route of
doing most of computation in large `t Hooft coupling limit and
extrapolate only at the end of the day. We expect that this strategy
works best when the quantities in question are not sensitive to the
`t Hooft coupling in the large $N_c$ limit, such as the chiral
coupling between pions and the baryon and the anomalous magnetic
moment of the nucleon.

Section 2 will review the holographic QCD model of Sakai and Sugimoto but
also serve as the reference section for many of the computations. We
will in particular introduce a conformal coordinate system which is
useful for dealing with the instanton soliton, which upon
quantization will be identified with baryons. In Section 3, we
consider basic static properties of baryons and include a careful
derivation of the size and the energetics. The instanton must then
be quantized to become physical baryon, and when the size is small,
namely when the 't Hooft coupling is very large, it can be treated
as a point-like object but with long range gauge field tails. The
resulting five-dimensional effective action with a novel and
essential magnetic coupling is derived in Section 4.

Beginning with Section 5, we start to discuss the chiral dynamics of
nucleons in four dimensions. We first describe how to reduce the
five-dimensional effective action of nucleons to four dimensions,
whose only nontrivial feature is a single magnetic coupling, and
produce a four-dimensional effective action of nucleons coupled to
the infinite tower of mesons. Some of the simplest predictions on
Yukawa coupling constants will be given as examples and compared to
experimental values. Section 6 will delve into numerical estimates
and extrapolation to realistic regimes, and points out subtleties
and potential problem in doing so.

Beginning with Section 7, we consider electromagnetic coupling of
nucleons. We review how the vector dominance in the meson sector
came about and show how this generalizes to nucleon sector rather
nontrivially. While the vector dominance here involves the entire
tower of vector mesons, we will show that truncating down to the
first four vector mesons, in both iso-scalar and iso-vector
sectors, respectively, provides a very good approximation to the
complete form factors of the model. As a bonus, one can also
compute the magnetic dipole moment of nucleons in Section 8 which
also compares favorably with experimental values. In Section 9, we
perform numeric analysis of electromagnetic form factors (Sachs
form factors) and also extract various nucleon charge radii. We
close with discussions.

An abbreviated version of this work has been reported elsewhere
\cite{Hong:2007kx} with emphasis on the derivation of the effective
action. The present paper expands upon the previous paper by
including more detailed derivation leading to the effective action
and exploring the implications comprehensively.

\section{A String Theory Model of Holographic QCD}

Among the holographic models of QCD proposed recently, one most
interesting and realistic model is the one by Sakai and Sugimoto
(SS)~\cite{sakai-sugimoto}, who considered  $N_c ~(\gg1)$ stack of
D4 branes and $N_F$ D8 branes in the background of Type
\uppercase\expandafter{\romannumeral 2}\,A superstring. The key
point of the model~\footnote{Unless otherwise stated, we follow the
notations of SS.} is that the flavor symmetries of the quark sector
are embedded into a $U(N_F)$ gauge symmetry in $R^{1+3}\times I$.
The fifth direction is topologically an interval, and the
four-dimensional low energy physics is found by restricting to the
modes that are localized near the ``origin" of this fifth direction.

The stack of D4 branes at low energy carries $SU(N_c)$ Yang-Mills
theory. In the large $N_c$ limit,
the dynamics of D4 is dual to a closed string theory in some
curved background with flux in accordance with the general AdS/CFT idea.
In the large 't~Hooft coupling limit, $\lambda\equiv g_{YM}^2N_c\gg1$,
and neglecting the gravitational back-reaction from the D8 branes, the
metric is \cite{Witten:1998zw}
\begin{equation}
ds^2=\left(\frac{U}{R}\right)^{3/2}\left(\eta_{\mu\nu}dx^{\mu}dx^{\nu}+f(U)d\tau^2\right)
+\left(\frac{R}{U}\right)^{3/2}\left(\frac{dU^2}{f(U)}+U^2d\Omega_4^2\right)
\end{equation}
with $R^3=\pi g_sN_cl_s^3$ and $f(U)=1-U_{KK}^3/U^3$. The
coordinate $\tau$ is compactified as $\tau=\tau+\delta\tau$ with
$\delta\tau=4\pi R^{3/2}/(3U_{KK}^{1/2})$.

\subsection{Five-Dimensional $U(N_F)$ Theory on D8-Branes }

The D8 branes, which share the coordinates $x^1,x^2,x^3$ with
the D4 branes, admit the massless quark degrees of freedom as
open strings attached  to both the D4 and D8 branes. The
effective action on a D8 brane, embedded in the D4
background, is the DBI action
\begin{equation} S_{D8}=-\mu_8 \int
{\rm
d}^9x\,e^{-\phi}\sqrt{-\det\left(g_{MN}+2\pi\alpha^{\prime}F_{MN}\right)}
+\mu_8\int\,\sum_p C_{p+1} \wedge \,e^{2\pi\alpha' F }\,,
\end{equation}
with
\begin{equation}
\mu_p=\frac{2\pi}{(2\pi l_s)^{p+1}} \: ,
\end{equation}
where $l_s^2=\alpha'$. $\sum C_{p+1}$ is a formal sum of the
antisymmetric Ramond-Ramond fields of odd-ranks, $C_1,C_3,C_5,C_7,
C_9$. These fields couple, respectively, to D0, D2, D4, D6, and D8
branes.

The D8 brane in this set-up occupies a $5D$ curved spacetime times
$S^4$ whose radius is position-dependent along $5D$.
The induced metric on D8 is
\begin{equation}
g_{8+1}=\left(\frac{U}{R}\right)^{3/2}\left(\eta_{\mu\nu}dx^{\mu}dx^{\nu}\right)
+\left(\frac{R}{U}\right)^{3/2}\left(\frac{dU^2}{f(U)}+U^2d\Omega_4^2\right) \: .
\end{equation}
We transform the coordinates so that the noncompact 5D part of the
metric is conformally flat,
\begin{equation}
g_{4+1}=H(w)\left(dw^2+\eta_{\mu\nu}dx^{\mu}dx^{\nu}\right) \:,
\end{equation}
where
\begin{equation}
w=\int_{U_{KK}}^U\frac{R^{3/2}dU^\prime}{\sqrt{{U^\prime}^3-U_{KK}^3}}\:.
\end{equation}
Note that the parameters of
dual QCD are mapped to the parameters here as
\begin{equation}
R^3=\frac{g_{YM}^2N_cl_s^2}{2M_{KK}}\:,\qquad U_{KK}=\frac{2g_{YM}^2N_cM_{KK}l_s^2}{9}\: ,
\end{equation}
where the KK mass $M_{KK}$ is the dimensionful free-parameter of the
theory. Note that
\begin{equation}
M_{KK}\equiv 3U_{KK}^{1/2}/2R^{3/2} \:.
\end{equation}
Another dimensionful quantity that appears in the chiral
Lagrangian formulation of QCD is $f_\pi$ which determines the
scale of chiral symmetry breaking. In terms of the above, we
have \cite{sakai-sugimoto,private}\footnote{See section 2.2.}
\begin{equation}
f_\pi^2=\frac{1}{54\pi^4}(g_{YM}^2N_c)N_c M_{KK}^2 \: .
\end{equation}
As was shown in detail by Sakai and Sugimoto, it is $M_{KK}$ that
enters the mass spectra of mesons. For real QCD, $M_{KK}$ would be
roughly $M_{KK}\sim m_N\sim 0.94$ GeV, while $f_\pi \sim$ 93 MeV, and
this requires
\begin{equation}
(g_{YM}^2N_c) N_c\sim 50 \: .
\end{equation}
For $N_c=3$, this gives
\begin{equation}
g_{YM}^2N_c\sim 17.
\end{equation}
This certainly is not big enough for truncating at the leading
order, indicating that it might be difficult to naively apply this
model to the realistic QCD regime. For this reason, the best we can
do is to look at dimensionless quantities in which the limiting
constants cancel out, such as ratio of masses of the mesons, and
hope that such quantities are insensitive to the precise values of
these physical parameters.

Note that this fifth coordinate is of finite range since
\begin{equation}
w_{max}=\int_0^\infty\frac{R^{3/2}dU}{\sqrt{U^3-U_{KK}^3}}
=\frac{1}{M_{KK}}\frac32\int_1^\infty\frac{d\tilde U}{\sqrt{\tilde U^3-1}}
\simeq \frac{3.64}{M_{KK}}<\infty \: .
\end{equation}
Thus, the 5D spacetime part of D8 brane is conformally equivalent to
an interval $[-w_{max}, w_{max}]$ times $R^{3+1}$. This makes the
search for smooth instanton solution rather subtle. This matter will
be discussed later in this paper. Another choice of coordinate
convenient for us is $z$ defined as
\begin{equation}
U^3=U_{KK}^3+U_{KK}z^2 \:  ,
\end{equation}
which is related to $w$ as
\begin{equation}
dw=\frac{R^{3/2}dU}{\sqrt{U^3-U_{KK}^3}}=
\frac{2R^{3/2}U_{KK}^{1/2}\:dz}{3(U_{KK}^3+U_{KK}z^2)^{2/3}}\:.
\end{equation}
Near origin $w\simeq 0$, we have the approximate relation,
\begin{equation}
M_{KK}w\simeq \frac{2}{3}\left(\frac{R}{U_{{KK}}}\right)^{3/2}\times(M_{KK}z)=\frac{z}{U_{KK}} \: ,
\end{equation}
which implies
\begin{equation}
U^3\simeq U_{KK}^3(1+M_{KK}^2w^2)
\end{equation}
for the conformally flat coordinate. This shows that the deviation
of the metric from the flat one is dictated entirely by the mass
scale $M_{KK}$. In fact the same is true of the full 10-dimensional
spacetime metric, and thus from this we can see that $M_{KK}$ is the
only mass scale of the theory in the low energy limit.

In the low energy limit, the worldvolume dynamics of the
D8 brane is well-described in terms of a derivative expansion
of the full stringy effective action. Extending this to multi-D8 brane
system gives the Yang-Mills action with a Chern-Simons term.
The Yang-Mills part of this effective action is
\begin{eqnarray}
&&\frac14\;\int_{8+1} \sqrt{-g_{8+1}}\;\frac{e^{-\Phi}}{{2\pi
(2\pi l_s)^5}}\; \tr F_{MN}F^{MN} \nonumber \\
&=&\frac14\;\int_{4+1}\sqrt{-g_{4+1}}
\;\frac{e^{-\Phi}V_{S^4}}{2\pi (2\pi l_s)^5} \;\tr F_{\hat m \hat
n}F^{\hat m\hat n} \: .\label{dbi}
\end{eqnarray}
Here $V_{S^4}$ is the position-dependent volume of the compact
part with
\begin{equation}
V_{S^4}=\frac{8\pi^2}{3}R^3U \: ,
\end{equation}
while the dilaton is
\begin{equation}
e^{-\Phi}=\frac{1}{g_s}\left(\frac{R}{U}\right)^{3/4} \:.
\end{equation}
The Chern-Simons coupling arises from the second set of terms
because $\int_{S^4}dC_3\neq 0$ takes a quantized value, and was
worked out by Sakai and Sugimoto in some detail. The answer after
integration over the four-sphere is
\begin{equation}
\frac{N_c}{24\pi^2}\int_{4+1}\omega_{5}(A)
\end{equation}
with $d\omega_5(A)=\tr F^3$.

\subsection{Chiral Lagrangian and Hidden Local Symmetry
(HLS)}\label{2.2}

The main point of this model is that the D8 comes with two
asymptotic regions (corresponding to UV) at $w\rightarrow\pm
w_{max}$ which are continuously connected via the infrared region
near $w=0$. The usual chiral symmetry $U(N_F)_L\times U(N_F)_R$
is implicitly embedded into the $U(N_F)$ gauge symmetry of D8
branes \cite{sakai-sugimoto}.  The $U(N_F)_{L,R}$ are the
remnant of the five-dimensional gauge symmetry; those on the
left-end and the right-end are each interpreted as $U(N_F)_{L,R}$,
respectively. While  the gauge symmetry is broken, its global
counterpart survives as $U(N_F)_{L+R}$.

The five-dimensional gauge field has
three polarizations. Thus the generic KK modes become massive vector
fields in four dimensions, namely massive vector mesons whose parity
is decided by the shape of the KK eigenfunction, while there is a
single massless adjoint multiplet which arises from the Wilson line
degrees of freedom, which are the pions. This can be seen more
clearly when one expands $A_\mu$ in terms of eigenmodes along $w$
directions, decomposing it into infinite towers of KK states as seen
by 4D observers. The lowest modes are then interpreted as the
low-lying vector mesons of the chiral Lagrangian formulation. In the
gauge $A_5=0$, this expansion was worked out by Sakai and Sugimoto.
Introducing a gauge function $\xi(x)$ at $w=0$, which is related to
the pion field in unitary gauge as
\begin{equation}
\xi^2(x)=U(x),\qquad U(x)=e^{2i\pi(x) /f_\pi} \: ,
\end{equation}
we have the following expansion\footnote{Our gauge field is defined
by $D=\partial-iA$, which differs from $D=\partial+A^{SS}$ of SS.},
\begin{equation}
A_\mu(x;w)= i\alpha_\mu(x)\psi_0(w)+i\beta_\mu(x) +\sum_n
a_\mu^{(n)}(x)\psi_{(n)}(w)
\end{equation}
with
\begin{equation}
\alpha_\mu(x)\equiv \{\xi^{-1},\partial_\mu\xi\}\simeq {2i\over
f_\pi}\partial_\mu\pi ,\qquad
\beta_\mu(x)\equiv\frac{1}{2}[\xi^{-1},\partial_\mu\xi]\simeq
{1\over 2f_{\pi}^2} [\pi, \partial_\mu\pi]\:,
\end{equation}
where $\psi_0(w)=\psi_0(w(z))={1\over\pi}\arctan\left(z\over
U_{KK}\right)$. Inserting this into the DBI-action (\ref{dbi}), we
can obtain a low-energy Lagrangian for the pions as well as massive
vector/axial-vector mesons.

As for the pions, this reproduces the Skyrme Lagrangian\footnote{After this
paper has appeared, we learned of a factor two error in Ref.~\cite{sakai-sugimoto}.
We thank S. Sugimoto for informing us \cite{private}.
In the present paper, all quantities are derived from the D-brane
physics and did not rely on the computations in
Ref.~\cite{sakai-sugimoto}. The only exception is the chiral Lagrangian
here, which affects the two coefficients $f_\pi^2$ for the
kinetic term and $1/e^2_{Skyrme}$ for the Skyrme term. This
enters physical quantities considered only indirectly
via the determination of $\lambda\sim 17 $ for the realistic QCD regime,
which affects slightly only the subleading corrections for
quantities we consider.}
\begin{equation}
{\cal L}_{pion}={f_\pi^2\over 4}\tr \left(U^{-1}\partial_\mu
U\right)^2 +{1\over 32 e^2_{Skyr
me}} \tr \left[ U^{-1}\partial_\mu U,
U^{-1} \partial_\nu U \right]^2
\end{equation}
with
\begin{equation}
f_\pi^2={1\over 54\pi^4}(g_{YM}^2 N_c) M_{KK}^2 N_c\:,\quad
e^2_{Skyrme}\simeq {54\pi^7\over 61} {1\over (g_{YM}^2 N_c)N_c}\:.
\end{equation}
For the massive tower of (axial) vector mesons, we have the standard
kinetic term
\begin{equation}
{\cal L}_{massive}=\sum_n{\rm tr}\left\{{1\over 2} F_{\mu\nu}^{(n)}
F^{\mu\nu(n)}+m_n^2 a_\mu^{(n)} a^{\mu(n)}\right\}\:,
\end{equation}
with $F^{(n)}_{\mu\nu}=\partial_\mu a^{(n)}_\nu-\partial_\nu
a^{(n)}_\mu$, plus various interactions between them as well as
with pions. When we decompose $U(N_F)$ into $SU(N_F)$ and $U(1)$,
the natural gauge generators are normalized as $\tr\, T^2=1/2$;
with $1/2$ in front of the trace we have the canonical normalization
for four-dimensional vector and axial-vector mesons.

The interesting point in this theory is that the gauge symmetry
localized in the fifth direction can be identified as an infinite
number of ``hidden local symmetries" (HLS) in four dimensions, and
each massive vector meson plays a role as a gauge field for some
part of them. A hidden gauge symmetry theory with the
$(\rho,\omega)$, the lowest members of the tower, was introduced
into hadron physics two decades ago by Bando et al~\cite{bandoetal}
and revived recently by Harada and Yamawaki~\cite{HY:PR}. The key
observation that led to the formulation of \cite{bandoetal} was that
the chiral field $U$ which figures in the low-energy dynamics of the
Goldstone pions possesses a hidden local symmetry that can be
exploited to bring the energy scale to $\sim 4\pi m_V/g$ (where
$m_V$ is the vector meson mass and $g$ is the hidden gauge
coupling). In the modern terminology, one can consider the hidden
gauge field so obtained as an emergent field as in other areas of
physics~\cite{georgi,polchinski}. (See Section 8 for more on this.)
In the current holographic model, this idea finds a natural home
simply because HLS arises automatically from the five dimensional
description which incorporates not only $(\rho,\omega)$ but the
entire tower of vector mesons. In our formulation we took a definite
gauge choice (i.e., unitary gauge) so that
$\xi=\xi_R=\xi_L^\dagger$. One can think of the SS model descending
top-down from string theory to the hidden local symmetry of QCD.
Indeed when restricted to the lowest member of the tower, the SS
action reduces to the HLS action of \cite{bandoetal} with $a=4/3$.

\section{Baryons as Small and Hairy Instantons \label{small}}

Conventional chiral Lagrangian approaches realize baryons as
Skyrmions, usually made of the pion field $U$ only. As we couple
higher massive vector mesons to the Skyrme action, the
size-stabilizing mechanism for topological solitons is significantly
affected by massive vector mesons. If we approach this problem from
the above five-dimensional viewpoint, however, it is natural to
consider the problem as a five-dimensional one. It has been known
for some time that what replaces the Skyrmion is the instanton
soliton since the two share the same topological winding number \cite{Atiyah:1989dq}.
However, what has not been clear is whether and how much of the
instanton is born out of the Skyrmion. As we will begin to see from
this section, the instanton interpretation of the baryon will give a
very different route to the low energy effective dynamics of the
baryons.

We know that
 a D4 brane wrapping the compact $S^4$ will correspond to a
baryon vertex on the $5D$ spacetime, which follows from an argument
originally given by Witten~\cite{witten-baryon}. On the D4-brane we
have a Chern-Simons coupling of the form,
\begin{equation}
\mu_4\int C_3\wedge 2\pi\alpha'{d\tilde A}=
2\pi\alpha'\mu_4\int dC_3\wedge \tilde A
\end{equation}
for D4 gauge field $\tilde A$. Since D4 wraps the $S^4$ which has
a quantized $N_c$ flux of $dC_3$, one finds that this term
induces $N_c$ unit of electric charge on the wrapped D4. The Gauss constraint
for $\tilde A$ demands that the net charge should be zero, however, and
the D4 can exist only if $N_c$ fundamental strings end on it. In
turn, the other end of fundamental strings must go somewhere, and
the only place it can go is D8 branes. Thus a D4 wrapping $S^4$
looks like an object with electric charge with respect to the
gauge field on D8. With respect to the overall $U(1)$ of the latter,
whose charge is the baryon number, the electric charge is $N_c$.
Thus,  we may identify the baryon as wrapped D4 with $N_c$
fundamental strings sticking onto it.

Of course, things are more complicated than this since D4 can dissolve into
D8 branes and become an instanton soliton on the latter. {}From D8's viewpoint,
a D4 wrapped on $S^4$ once is interchangeable with the unit instanton
\begin{equation}
\frac{1}{8\pi^2}\int_{R^3\times I} \tr F\wedge F=1 \: ,
\end{equation}
as far as the conserved charge goes. This follows from a Chern-Simons term
on D8,\footnote{Recently, this term was shown to play an interesting role
in a different aspect of baryonic physics with finite baryon density
\cite{Domokos:2007kt}.}
\begin{equation}
\mu_8\int_{R^{3+1}\times I \times S^4}\, C_{5} \wedge 2\pi^2(\alpha')^2
\tr F\wedge F =\mu_4 \int_{R^{0+1}\times S^4}
C_5\wedge\frac{1}{8\pi^2}\int_{R^3\times I}\,\tr F\wedge F \:,
\end{equation}
which shows that a unit instanton couples to $C_5$ minimally, and
carries exactly one unit of D4 charge. When the size of the
instanton becomes infinitesimal, it can be freed from D8's, and
this is precisely D4. From the viewpoint of D4, this corresponds
to going from the Higgs phase into the Coulomb phase.

In flat background geometry and no flux, the moduli space of D4
contains both the Coulomb branch where D4 maintains its identity
separated from D8, and the Higgs branch where D4 is turned into a
finite size Yang-Mills instanton on D8. With the
present curved geometry, this is no longer a matter of choice. The
energy of the D4 will differ depending on the
configurations. As we will see shortly, to the leading
approximation, the D4 will settle at the border of the two branches,
both of which disappear apart from basic translational degrees of
freedom along $R^{3+1}$. The reason for why D4 cannot dissociate
away from D8 is obvious. The D4 has $N_c$ fundamental strings
attached, whose other ends are tied to D8. Moving away from D8 by
distance $L$ means acquiring extra mass of order $N_cL/l_s^2$ due to
the increased length of the strings, so the D4 would stay on top of
D8 for a simple energetics reason.
The question is then how small or big will a D4 spread inside D8 as
an instanton. Consider the kinetic part of $D8$ brane action,
compactified on $S^4$, in the Yang-Mills approximation,
\begin{equation}
-\frac14\;\int\sqrt{-g_{4+1}} \;\frac{e^{-\Phi}V_{S^4}}{2\pi (2\pi
l_s)^5} \;\tr F_{\hat m\hat n}F^{\hat m\hat n}\:.
\end{equation}
After taking the volume of $S^4$, the dilaton, and the conformally
flat metric, this reduces to
\begin{equation}
-\;\int dx^4 dw
\;\frac{1}{4e^2(w)} \;\tr F_{mn}F^{mn}\:,
\end{equation}
where the contraction is with respect to the flat metric $dx_\mu dx^\mu+dw^2$
and the position-dependent electric coupling  $e(w)$ of this five dimensional
Yang-Mills is such that
\begin{equation}
\frac{1}{e^2(w)}\equiv \frac{8\pi^2 R^3U(w)}{3 (2\pi l_s)^5 (2\pi
g_s)}\:.
\end{equation}
In the SS model, the string coupling $g_s$ is related to the
dimensionful parameters and four-dimensional Yang-Mills coupling of
the QCD as,
\begin{equation}
2\pi g_s=\frac{g_{YM}^2}{M_{KK}l_s}\:,
\end{equation}
so we find
\begin{equation}
\frac{1}{e^2(w)}
=\frac{(g_{YM}^2N_c)N_c}{108\pi^3}M_{KK}\frac{U(w)}{U_{KK}}
\end{equation}
Since an instanton has
\begin{equation}
\int \tr F_{mn}F^{mn}=2\int \tr F\wedge F=16\pi^2\:,
\end{equation}
a point-like instanton that is localized at $w=0$ would
have the energy
\begin{equation}
m_B^{(0)}\equiv \frac{4\pi^2}{e^2(0)}=\frac{(g_{YM}^2N_c)N_c}{27\pi}M_{KK}\:.
\end{equation}
This mass also equals that of an $S^4$ wrapped D4 located at $w=0$,
in accordance with the string theory picture of the instanton
\cite{sakai-sugimoto}. If
the instanton gets bigger, on the other hand, the configuration
costs more and more energy, since $1/e^2(w)$ is an increasing
function of $|w|$, thus the leading behavior of the instanton is to
collapse to a point-like instanton.

However, the $N_c$ fundamental strings attached to D4
manifest themselves as $N_c$ units of electric charge
on D8's. There will be in general Coulomb repulsion
among these electric charges, and this would favor spreading
of instanton to a finite size. So it is the competition
of the two effects, mass of instanton vs. Coulomb energy
of fundamental strings. For very small instanton of size
$\rho$, the energy picks up a size-dependent piece from
the action of Yang-Mills field which goes as
\begin{equation}\label{mass}
\sim \frac16\, m_B^{(0)}M_{KK}^2\rho^2\:,
\end{equation}
while the five dimensional Coulomb energy goes as
\begin{equation}\label{C}
\sim \frac12\times \frac{e(0)^2N_c^2}{10\pi^2\rho^2}\:,
\end{equation}
provided that $\rho M_{KK} \ll 1$. The estimate of energy
here takes into account the spread of the instanton density
$D(x^i,w)\sim \rho^4/(r^2+w^2+\rho^2)^4$, but ignores
the deviation from the flat geometry along the four
spatial directions.

We kept an overall factor of $1/2$ in the Coulomb energy separated
from the rest because it deserves a further explanation. The rest of
the term is  the five dimensional $U(1)$ (with electric coupling
constant $e(0)$) Coulomb energy for charge  $N_c$ whose distribution
follows the instanton density $D(x^i,w)$. To see the origin of the
additional factor of $1/2$, recall that the Chern-Simons term
responsible for this charge is
\begin{equation}
\frac{N_c}{24\pi^2}\int \tr \left(A \wedge F\wedge F +\cdots \right)\:,
\end{equation}
from which we obtain the coupling between instanton  $\bar F$ and
the rest of the gauge field as
\begin{equation}
\frac{N_c}{8\pi^2}\int \tr \left(A \wedge  \bar F\wedge \bar F  \right)\:.
\end{equation}
Gauge rotating a single instanton into the form
\begin{equation}\label{charge}
\frac{N_c}{8\pi^2}\,\bar F\wedge \bar F=\frac{N_c D(x^i,w)}{2}
\left(\begin{array}{ccccc}
1&0 &0&\cdots & 0 \\
0&1 &0&\cdots &0\\
0&0 &0&\cdots &0 \\
\cdots &\cdots &\cdots & \cdots & \cdots \\
0&0&0&\cdots&0
\end{array}\right)dx^3\wedge dw\:,
\end{equation}
we have a minimal coupling to the instanton worldline
\begin{equation}
N_c \sum_a \int A^a\, \tr \left(T^a I_2/2
 \right)\:,
\end{equation}
where $I_2$ denotes the matrix in Eq.~(\ref{charge}).

For $N_F=2$,  only the
trace part of $A$ can couple to $I_2$, and
$A_{U(1)}$ in
\begin{equation}
A=A_{U(1)}\left(\begin{array}{cc} 1&0\\0&1\end{array}\right)
\end{equation}
sees charge $N_c$ on top of a single instanton. However, the kinetic
term for $A_{U(1)}$ would have the coefficient $1/2e^2$ instead of
$1/4e^2$, which  changes the effective electric coupling constant
and introduces a factor of $1/2$ to the Coulomb energy. For $N_F>2$,
the same factor of $1/2$ arises for more complicated reasons. There
are now $N_F-2$ vector fields in $SU(N_F)$, as well as the $U(1)$
vector field from the trace part, that couple to this charge under
the above Chern-Simons term. Each of them contributes some fraction
of the above $U(1)$ Coulomb energy. But the sum can be seen to be
always 1/2.

More simply, this reduction can be seen from the fact that the total
electric charge $N_c$ on the instanton is shared, evenly split,
 by a pair of mutually orthogonal $U(1)$'s of $U(N_F)$,
which is evident in the form of $I_2$. In each sector the
electric charge generates the Coulomb energy, proportional
to $(N_c/2)^2$. Since the total Coulomb energy is obtained by a
sum, we find $2\times (N_c/2)^2=N_c^2/2$ in place of $N_c^2$.

The size of the small instanton is determined where the combined
energy is minimized \cite{Hong:2007kx,Hata:2007mb}\footnote{The derivation of
soliton size in this paper is an expanded version of that in Ref.~\cite{Hong:2007kx}. Note that an
independent derivation was given in Ref.~\cite{Hata:2007mb} which appeared simultaneously with the former.}
\begin{equation}\label{size}
\rho^2_{baryon}\simeq \frac{1}{M_{KK}}\sqrt{\frac{3e(0)^2N_c^2}{10 \cdot \pi^2 m_B^{(0)}}}
=\frac{\sqrt{2\cdot 3^7\cdot\pi^2/5}}{M_{KK}^2(g_{YM}^2N_c)}\:,
\end{equation}
and
\begin{equation}
\rho_{baryon}\sim \frac{9.6}{M_{KK}\sqrt{g_{YM}^2N}} \:.
\end{equation}
For an arbitrarily large 't Hooft coupling limit, the size of
baryon is then significantly smaller than the scale of the dual
QCD. Subsequently the mass correction to the baryon due to its
5-dimensional electric coupling
\begin{equation}
m_0^e \simeq
\frac{1}{3}m_B^{(0)}(M_{KK}\rho_{baryon})^2 \simeq
\frac{31 }{g_{YM}^2N_c} m_B^{(0)} \ll m_B^{(0)}\label{corr}
\end{equation}
is also small if the  't Hooft coupling is arbitrarily large.

In the next section, we will thus assume a point-like baryon as a
leading approximation and incorporate baryons into the chiral
Lagrangian formulation. While this is a meaningful computation in
holographic QCD setting, matching the scales and couplings to the realistic
QCD requires a further refinement, since as mentioned, the scales
$M_{KK}$ and $f_\pi$ are actually too low to insist on very large
value of $g_{YM}^2N_c$.

In making the estimate above,  we ignored so far details of the
geometry away from the origin. For instance, the spatial part of the
geometry is conformally equivalent to $R^3\times I$, instead of
$R^4$. It is unlikely that
the lowest energy configuration is a self-dual instanton solution
based on $R^4$, yet we use it as a trial configuration to estimate
the potential. We believe that this will not affect the asymptotic
estimate in this section, when the size of the instanton is very
small compared to the effective length of the fifth direction $\sim
1/M_{KK}$. This subtlety would be more important for larger
instanton size, as we will discuss in Section.~\ref{fat}.

\section{A 5D Effective Field Theory of the Baryon \label{5DB}}

We saw in the previous section that in the large 't Hooft coupling
limit, the underlying instanton configuration for the baryon is
rather small. Since the instanton is a small object in 5D sense, we
may treat it as a point-like quantum field in 5D in a natural
way. Upon quantizing the collective coordinates of the solitonic
configuration, there are a variety of baryonic excitations with
different spins and flavor charges. From the gauge theory point of
view, baryons are composed of $N_c$ elementary quarks forming a
color singlet through a total anti-symmetrization of their color
indices. The remaining spin and flavor indices together must then
form a totally symmetric combination. It is always possible to have
one such combination via totally anti-symmetrizing both spin and
flavors, giving us the minimal spin and flavor quantum numbers. For
even $N_c$ we would have a spin 0 baryon, while a fermionic spin
$1\over 2$ baryon would occur when $N_c$ is odd. Having in mind an
extrapolation to the real QCD, we restrict ourselves to the case of
fermionic baryons, and the effective field $\cal B$ would mean a 5D
Dirac spinor field. For simplicity we will consider $N_F=2$ and
consider the lowest baryons which form the proton-neutron doublet
under $SU(N_F=2)$. We are thus lead to introduce an isospin 1/2
Dirac field ${\cal B}$ for the five-dimensional baryon.

{}From the invariance under local coordinate as well as local gauge
symmetries on the D8 branes reduced along internal $S^4$, the
leading 5D kinetic term for $\cal B$ is simply the standard Dirac
kinetic term in the curved space in addition to a position
dependent mass term that we will specify shortly,
\begin{equation}
-\int\,dz \int\,dx^4\, \left[i\bar{\cal B} \Gamma^{\hat m} D_{\hat m} {\cal
B} +i m_b(z)\bar{\cal B} {\cal B}\right]\:, \label{5Dbaryon}
\end{equation}
where $D_m=\partial_{\hat m}+{1\over 4}\Gamma_{\hat n\hat p}
\omega_{\hat m}^{\hat n\hat p}+i A_{\hat m}^a
T^a$ with $T^a$ a representation matrix for $\cal B$. To determine
$m_b(z)$, it is convenient to work in the conformally flat
coordinate $(w,x^\mu)$ where the spin connection piece can be
removed upon suitable rescaling of the $\cal B$ field \cite{Hong:2006ta}.
Thus we have
\begin{equation}
-\int dw\int d^4 x\,\left[i\bar{\cal
B}\gamma^\mu(\partial_\mu-i A_\mu^a T^a){\cal B}+i\bar {\cal
B}\gamma^5\partial_w {\cal B} +i m_b(w)\bar{\cal B}{\cal
B}\right]\:,\label{5dfermion}
\end{equation}
in the conformal coordinate system and with the $A_5=0$ gauge.
Here, $\gamma^{\mu}$ and $\gamma^5$ are the
standard gamma matrices in the flat space.

The position-dependent mass term requires a further clarification.
An elementary excitation approximately
localized at the position $w$ would have an energy $m_b(w)$, which
must be identified as the energy of an $S^4$-wrapped D4 brane
localized at the position $w$. From the DBI action of D4 brane,
this mass is found to be
\begin{equation}
m_B^{(0)}\cdot \left(U\over U_{KK}\right)\:,
\end{equation}
where $U$ should be considered as an implicit function of $w$ upon
the coordinate change from $U$ to $w$, and
\begin{equation}
m_B^{(0)}={(g_{YM}^2 N_c )\cdot N_c \over 27\pi}\cdot
M_{KK}={\lambda N_c\over 27\pi}\cdot M_{KK}\:,
\end{equation}
with $M_{KK}={3\over 2}\left(U_{KK}\over R\right)^{3\over
2}U_{KK}^{-1}$. In addition, there is a self-energy $m_0^e$ coming
from the 5D $U(1)$ field which stabilizes the instanton at some
small but finite size. Since this self-energy is a local effect as
the baryon size is negligible, this effect should be, at least
approximately, independent of the position $w$ and the resulting
$m_b(w)$ will be
\begin{equation}
m_b(w)=m_B^{(0)}\cdot \left(U\over U_{KK}\right)+m_0^e\:.
\end{equation}
In large 't Hooft coupling limit, the estimate (\ref{corr}) shows
that the Coulomb energy $m_0^e$ is negligible compared to the first
piece. But, we will keep it in our later numerical analysis for
completeness.\footnote{One may also worry about self-energy from
$SU(N_F=2)$ gauge field on D4 branes. However, $m_0^e$ scales
linearly with $N_c$ because the baryon has charge $N_c$ with respect
to $U(1)_V$, while there is no such scaling for $SU(N_F=2)$. In the
present model, $m_0^e$ is suppressed due to the further requirement
of large 't Hooft coupling.}

However, this cannot be the complete form of the baryon action. As
we saw above the baryon is represented by a small instanton soliton,
which comes with a long range tail of the gauge field of type
$F\sim \rho_{baryon}^2/r^4$.
Since we are effectively replacing the baryon by a point-like field
${\cal B}$, there should be a coupling between a ${\cal B}$ bilinear
and the five-dimensional gauge field such that each ${\cal
B}$-particle generates such a long range tail on $F$.\footnote{The
same type of consideration was employed by Adkins, Nappi and Witten
(ANW)~\cite{ANW} to compute $g_{\pi NN}$ which is related to $g_A$
by Goldberger-Treiman relationship. }
The minimal coupling originates from fundamental strings attached to
D4, and reflects the fact that the instanton carries additional
electric charge. This coupling cannot generate a self-dual or
anti-self-dual configuration.

As we will see shortly, there is only one vertex that can
reproduce the right long-range tail. In our conformal coordinate
$(x^\mu,w)$, the action including the gauge field and the baryon
field must read as\footnote{As usual, we define $\gamma^{mn}=[\gamma^m,\gamma^n]/2$.}
\begin{eqnarray}
&&\int d^4 x dw\left[-i\bar{\cal B}\gamma^m D_m {\cal B}
-i m_b(w)\bar{\cal B}{\cal B} +g_5(w){\rho_{baryon}^2\over
e^2(w)}\bar{\cal B}\gamma^{mn}F_{mn}{\cal B} \right]\nn
&-&\int d^4 x  dw {1\over 4 e^2(w)} \;\tr\, F_{mn}F^{mn}\,,
\label{5dfermion1}
\end{eqnarray}
where $\rho_{baryon}$ is the stabilized size of the 5D instanton
representing baryon, and $g_5(w)$ is an unknown function whose
value at $w=0$ can be determined as follows. Throughout this article
we will refer to the last term in the first line as the magnetic coupling.
Its coefficient function
is displayed in a particular form with the known function $\rho_{baryon}^2/e(w)^2$
factored out. This is done for the sake of later convenience, where we
compute $g_5(0)$ which turns out to be of a purely geometrical origin.

The uniqueness of the operator can be seen from the
long range behavior of the instanton. The field strength decays as $1/r^4$,
which in five dimensions is one power higher than the Coulomb field.
This requires dimension six operators (i.e. one dimension higher than
the kinetic term)
which contain a cubic term with a baryon bilinear current and the $SU(N_F)$ gauge field.
These requirements, together with the approximate Lorentz symmetry,
pick out the above form of the operator uniquely. The only other
choice would be $\bar{\cal B}(*F)_{mnk}\gamma^{mnk}{\cal B}$, but this
is actually equivalent to the above, thanks to the five-dimensional
Clifford algebra. Given that this operator is the unique possibility, the
remaining question is whether this operator is really capable of
the task in hand and if so how to derive the coupling strength $g_5$,
to which we devote the rest of the section.

The instanton must be located at $w=0$ along the fifth direction,
and generates a source term to Yang-Mills field $F_{MN}$. Provided
that the instanton size, $\rho_{baryon}$ is small enough, we only
need to consider the immediate vicinity of $w=0$ where the geometry
is $R^{4+1}$ approximately. Take the 5-dimensional Dirac matrices of
the form
\begin{equation}
\gamma^0=\left(\begin{array}{rr} 0 & -1 \\ 1 & 0\end{array}\right),\quad
\gamma^i=\left(\begin{array}{cc} 0 & \sigma_i \\ \sigma_i & 0\end{array}\right),\quad
\gamma^5=\left(\begin{array}{rr} 1 & 0 \\ 0 & -1\end{array}\right).
\end{equation}
The on-shell condition of the baryon field is then
\begin{equation}
\left(\begin{array}{cc}i\partial_5 & -i\partial_t+i\sigma^i \partial_i  \\
i\partial_t+i\sigma^i\partial_i  & -i\partial_5\end{array}\right){\cal B}=-im_b{\cal B}\:,
\end{equation}
which can be solved by writing the upper 2-component part
of ${\cal B}$ as ${\cal U}\,e^{-iEt+i\vec p\cdot\vec x}$,
and approximating $m_b$ by its central value,
\begin{equation}
{\cal B}=\left(\begin{array}{c}{\cal U}\\ \frac{E-\sigma\cdot p}{-im_b-p_5}{\cal U}\end{array}
\right)e^{-iEt+i\vec p\cdot\vec x}
\quad\rightarrow\quad
{\cal B}=\left(\begin{array}{r}{\cal U}\\ \pm i {\cal U}\end{array}
\right)e^{\mp im_bt}
\end{equation}
for general plane-wave and for the $p=0$ limit. The two signs
originate from the sign of $E/m_b$ and thus correspond to the baryon
and the anti-baryon, respectively.

This spinor configuration sources the Yang-Mills field since\footnote{Note that
\begin{equation}
\gamma^0\gamma^{jk}=\left(\begin{array}{cc} 0  & -i\epsilon^{jki}\sigma_{i} \\ i\epsilon^{jki}\sigma_{i}  & 0 \end{array}\right),\quad
\gamma^0\gamma^{5i}=\left(\begin{array}{cc} \sigma_i  &  0\\ 0 & \sigma_i\end{array}\right),\quad
\gamma^0\gamma^{0m}=-\gamma^m\: .
\end{equation}}
\begin{equation}
\bar{\cal B}\gamma^{mn}F_{mn}{\cal B}\quad\rightarrow\quad
\pm F_{jk}^a\left[ {\cal U}^\dagger\tau^a\epsilon^{jki}\sigma_i{\cal U}\right]
+2F_{5i}^a\left[{\cal U}^\dagger\tau^a\sigma_i{\cal U}\right] \:,
\end{equation}
where we assumed a gauge-doublet under $SU(N_F=2)$ with $2\times 2$
generators $(\tau^a/2)_{AB}$. Terms linear in $F_{0M}$ vanish
identically when $p=0=p_5$, thanks to the on-shell condition.

Note that the proper normalization of Dirac spinor demands
$2 {\cal U}^*_{\alpha A}{\cal U}^{\alpha A}=1$. Defining the bilinear
\begin{equation}
\langle \sigma_i\tau^a\rangle_{\cal B}=2\left[{\cal U}^\dagger\sigma_i\tau^a{\cal U}\right] \:,
\end{equation}
we thus find
\begin{equation}\label{sd}
\bar{\cal B}\gamma^{mn}F_{mn}{\cal B}\quad\rightarrow\quad
\pm \frac12 F_{jk}^a \epsilon^{jki}\langle \sigma_i\tau^a\rangle_{\cal B}
+F_{5i}^a\langle \sigma_i\tau^a\rangle_{\cal B} \:.
\end{equation}
Clearly the spinor bilinear couples to self-dual or an anti-self-dual
part of the gauge field strength, regardless of the detailed values of
$\langle \sigma_i\tau^a\rangle_{\cal B}$. Thus, if we relate the
effect of the latter to the smearing of the classical long-range field
due to quantization of the instanton, identification of ${\cal B}$ as the
effective field for isospin 1/2 baryons would be complete and this would
give information on the coupling strength $g_5$.

However, for clarity, let us first try to search for an instanton-like long-range field.
For instance, one choice for ${\cal U}$ that generates instanton-like
field is a spin-isospin locked state of the form,
\begin{equation}
{\cal U}_{\alpha A}=\frac{i}{2} \epsilon_{\alpha A} \:,
\end{equation}
in which case $\langle \sigma_i\tau^a\rangle_{\cal B}=-\delta_i^a$
so that the source term (with the upper sign) is  $-F_{mn}^a\bar \eta^a_{mn}/2$
with the anti-self-dual 't Hooft symbol
$\bar \eta$ ($m,n=1,2,3,5$ and $a=1,2,3$) \cite{'tHooft:1976fv}.
Now assume that such a source appears in a localized form at the
origin. The gauge field far away from the source obeys (in an
appropriate gauge)
\begin{equation}
\nabla^2A_m^a= 2g_5(0)\rho^2_{baryon}\bar\eta^a_{mn} \partial_n \delta^{(4)}(x) \:,
\end{equation}
whose solution goes as
\begin{equation}
A^a_m= -\frac{g_5(0)\rho^2_{baryon}}{2\pi^2}\bar\eta^a_{mn} \partial_{n}\frac{1}{r^2+w^2} \:.
\end{equation}
The general shape of the long-range field is consistent with the identification
of the baryon as the instanton. Since the actual instanton solution in 't Hooft ansatz
has \cite{Jackiw:1976fs}
\begin{equation}
A^a_m =  -\bar\eta^a_{mn}\partial_n \log\left(1+\frac{\rho^2}{r^2+w^2}\right)
\simeq -\rho^2 \bar\eta^a_{mn}\partial_n \frac{1}{r^2+w^2}\:,
\end{equation}
one may be tempted to fix $g_5(0)$ as $2\pi^2$.

However, the right prescription is to match the states in ${\cal B}$
with {\it quantized} instanton. An $SU(2)$ instanton of a fixed size
has three gauge collective coordinates, spanning $SU(2)/Z_2$, which
can be represented by a special unitary matrix $S$ of size $2\times
2$. Quantization of $S$ can lead to spin 1/2 and isospin 1/2 states
with proper choice of boundary condition on $S^3/Z_2=SU(2)/Z_2$.
This part of story proceeds identically with that of Skyrmions in 4
dimensions, which was explained in much detail by Adkins, Nappi and
Witten~\cite{ANW}.

One consequence of this quantization procedure is that the long range
field of the instanton is modified due to quantum fluctuation
of instanton along different global gauge directions. While the
classical solution has
\begin{equation}
S^\dagger A^a_M  \frac{\tau^a}{2}S =\sum_b  A^a_M \frac{\tau^b}{2}\;\left(
\tr\left[S^\dagger \frac{\tau^a}{2}S\tau^b \right]\right)
\end{equation}
for some arbitrary but fixed choice of the unitary matrix $S$, the
quantum consideration replaces the classical coefficients by
expectation values
\begin{equation}
\left(
\tr\left[S^\dagger \frac{\tau^a}{2}S\tau^b \right]\right)
\quad\Rightarrow\quad \left\langle
\tr\left[S^\dagger \frac{\tau^a}{2}S\tau^b \right]\right\rangle \:,
\end{equation}
which effectively lessen the strength of long-range gauge field.
We may identify the states contained in ${\cal B}$ as spin 1/2
and isospin 1/2 wavefunctions of the instanton, in which case
there is an identity,
\begin{equation}\label{anw}
\left\langle
\tr\left[S^\dagger \frac{\tau^a}{2}S\tau^b \right]\right\rangle_{\cal B}
=-\frac{1}{3}\langle \sigma_b\tau^a\rangle_{\cal B} \:.
\end{equation}
This can be seen by an explicit quantization, which is
mathematically identical to the one used by ANW~\cite{ANW} on the
Skyrmion case.

Specializing back to the case of ${\cal U}=i\epsilon$, where
$\langle \sigma_i\tau^a\rangle_{\cal B}=-\delta_i^a$,
note that the classical counterpart would have corresponded to
the choice $S=1$ so that
\begin{equation}
\left(\tr\left[S^\dagger \frac{\tau^a}{2}S\tau^b \right]\right)\biggr\vert_{S=1}=\delta^b_a \:,
\end{equation}
while the actual comparison has to be made with its quantum counterpart
\begin{equation}
\left\langle\tr\left[S^\dagger \frac{\tau^a}{2}S\tau^b \right]\right\rangle_{\cal B}
=-\frac{1}{3}\langle \sigma_b\tau^a\rangle_{\cal B}=\frac13\delta_a^b \:.
\end{equation}
This tells us that when making comparison between the long range
part of quantized instanton solution, and the long range field generated
by the baryon source, we must include a factor of 1/3 on the instanton
size. Thus, we conclude that
\begin{equation}
g_5(0)=\frac{2\pi^2}{3} \:.
\end{equation}
This fixes the value of $g_5(w)$ at origin of the fifth direction.
Finding the form of the function $g_5(w)$ for general value of $w$
seems very difficult from the present approach. However, for very
small size of baryon/instanton, which is guaranteed by a large 't
Hooft coupling, $\lambda=g_{YM}^2N_c$, only the central value will
enter the physics and corrections are suppressed by inverse powers
of $\lambda$.

In the above, we have extracted $g_5(0)$ by comparing the quantized
instanton and the spinor state for a particular spin-isospin locked
state. For a complete check, we must consider more general
states with spin 1/2 and isospin 1/2, for which it suffices to
rewrite Eq.~(\ref{sd}), say, with the upper sign choice, as
\begin{eqnarray}
&&\frac12 F_{jk}^a \epsilon^{jki}\langle \sigma_i\tau^a\rangle_{\cal B}
+F_{5i}^a\langle \sigma_i\tau^a\rangle_{\cal B}=
\frac{1}{2}F_{jk}^a\bar\eta^b_{jk}\langle \sigma_b\tau^a\rangle_{\cal B}
+F_{5k}^a\bar\eta^b_{5k}\langle \sigma_b\tau^a\rangle_{\cal B}\nonumber\\
\nonumber\\
&=&\frac{1}{2}F_{mn}^a\bar\eta^b_{mn}\langle \sigma_b\tau^a\rangle_{\cal B}
=-\frac{3}{2}F_{mn}^a\bar\eta^b_{mn}
\left\langle\tr\left[S^\dagger \frac{\tau^a}{2}S\tau^b \right]\right\rangle_{\cal B} \:,
\end{eqnarray}
where the last step used the identity Eq.~(\ref{anw}) between expectation
values in two different description. Since $S$ represents $SU(N_F=2)$
rotation on the soliton side of the picture, the long range field
generated by such a source would mimic that of the instanton field
expectation value, evaluated on arbitrary quantized instanton state
with spin 1/2 and isospin 1/2. We can follow a similar procedure above
for the spin-isospin locked state, which shows that  the on-shell
degrees of freedom of ${\cal B}$ can be matched with the spin 1/2
isospin 1/2 sector states of the quantized instanton, given the effective
action for ${\cal B}$ and $g_5(0)=2\pi^2/3$.

\section{The Chiral Dynamics of the Nucleons in Four Dimensions}

In the current effective theory approach, the physical 4D nucleons
would arise as the lowest eigenmodes of this 5D baryon along $w$
coordinate, which should be a mode localized near $w=0$. From the
string theory picture where the solitonic configuration for a
baryon comes from a melted D4 brane inside $N_F=2$ D8 branes,
there are $N_c$ fundamental strings ending on the D8 branes out of
the D4 brane, which are nothing but the elementary quarks in the
gauge theory view point. In the limit of large $\lambda$, this
consideration leads us to treat the five-dimensional baryon as a
point-like object in the doublet representation under $SU(N_F=2)$
with the effective action in (\ref{5dfermion1}).
While the generalization to excited baryons, such as $\Delta$'s,
should be straightforward, we will consider isospin doublets in
this work. In particular, the lowest-mass eigenstates in
4D sense are nothing but the nucleons (protons and neutrons),
whose low energy dynamics will be explored for the rest of the paper.

What we need now
is to reduce this five-dimensional action down to four dimensions
and extract the couplings between the nucleon and the infinite
tower of mesons.
In the usual chiral Lagrangian approach of QCD, the nucleon is
often treated as a point-like Dirac field $B$, just as in our
five-dimensional approach. In doing so, the form factors of the
nucleons would be then encoded in how the nucleons couples to pions
and all the massive vector mesons. The leading quadratic part of
the nucleon effective action one usually writes down is
\begin{equation}
\int dt dx^3 \;{\cal L}_4=-\int dt dx^3 \; \bar B(i\gamma^\mu
D_\mu+im_B+g_A\gamma^\mu\gamma^5A_\mu)B +\cdots \:,
\end{equation}
where the covariant derivative
\begin{equation}
D_\mu=\partial_\mu -i V_\mu
\end{equation}
encodes the coupling to the massive vector meson $v_\mu$
\begin{equation}
V_\mu=v_\mu+i\beta_\mu(x)=v_\mu+\frac{i}{2}[\xi^{-1},\partial_\mu\xi]
\end{equation}
in a manner consistent with the hidden local gauge symmetry. The
axial coupling provides the simplest vertex of this theory whereby
nucleon emits a single pion. In terms of $\xi$, we have
\begin{equation}
A_\mu=\frac{i}{2}\,\alpha_\mu \simeq -\frac{1}{f_\pi}\,\partial_\mu \pi +O(\pi^3) \:.
\end{equation}
The goal of this section is to reproduce this structure and more
from our five-dimensional effective action.

\subsection{4D Nucleons and Dimensional Reduction}

To make the preceding discussion concrete, let us perform KK-mode
expansion for the action (\ref{5dfermion}) to obtain the spectrum of
spin-$1\over 2$ baryons in the large $\lambda N_c=(g_{YM}^2N_c)N_c$
limit. The lowest state is identified as the nucleon. The gauge
field $A_\mu$ on the $N_F=2$ D8 branes also has a mode expansion,
including pions and $\rho$ mesons, that is discussed in the
preceding sections. From these we can read off the couplings of
nucleons to mesons via numerical analysis.

We mode expand ${\cal B}_{L,R}(x^\mu,w)=B_{L,R}(x^\mu)f_{L,R}(w)$
where $\gamma^5 B_{L,R}=\pm B_{L,R}$ are 4D chiral components,
with the profile functions $f_{L,R}(w)$ satisfying
\begin{eqnarray}
\partial_w f_L(w)+m_b(w) f_L(w) &=& m_B f_R(w)\:,\nonumber\\
-\partial_w f_R(w)+m_b(w) f_R(w) &=& m_B f_L(w)\:,
\end{eqnarray}
in the range $w\in[-w_{max},w_{max}]$. The 4D Dirac field for
the nucleon is then reconstructed as
\begin{equation}
B=\left(\begin{array}{c} B_L \\ B_R \end{array}\right)\:,
\end{equation}
and the eigenvalue $m_B$ is the mass of the nucleon mode $B(x)$.

The eigenfunctions $f_{L,R}(w)$ are also normalized to unit norm
\begin{equation}
\int_{-w_{max}}^{w_{max}} dw\,\left|f_L(w)\right|^2 =
\int_{-w_{max}}^{w_{max}} dw\,\left|f_R(w)\right|^2 =1\:,
\end{equation}
for $B(x)$ to have the standard 4D kinetic term. As we approach $w\to \pm w_{max}$,
$m_b(w)$ diverges as,
\begin{equation}
m_b(w) \sim {1\over (w\mp w_{max})^2}
\end{equation}
and the above equations have normalizable eigen-functions with a
discrete spectrum of $m_B$. It is more convenient to consider a
second-order equation for $f_{L,R}(w)$
\begin{eqnarray}
&&\left[-\partial^2_w -\partial_w m_b(w)+(m_b(w))^2\right]
f_L(w)=m_B^2 f_L(w)\:,\nonumber \\
&&\left[-\partial^2_w +\partial_w m_b(w)+(m_b(w))^2\right]
f_R(w)=m_B^2 f_R(w)\:.\label{eigeneq}
\end{eqnarray}
Note that there is  a 1-1 mapping of eigenmodes
with $f_R(w)=\pm f_L(-w)$. Due to the
asymmetry under $w\to -w$ in the term $-\partial_w m_b(w)$ above,
$f_L(w)$ tends to shift to the positive $w$ side, and the opposite
happens for $f_R(w)$. This will then give us a non-vanishing
contribution to the axial coupling of the nucleon to the pions, as
we will see shortly.

The gauge field $A_\mu$, in the $A_5=0$ gauge, has a mode
expansion
\begin{equation}
A_\mu(x,w)=i\alpha_\mu(x)\psi_0(w) +i\beta_\mu(x)+\sum_n
a_\mu^{(n)}(x)\psi_{(n)}(w)\:,
\end{equation}
where $\hat\Psi_0(z)\equiv \psi_0(w(z))={1\over\pi}\arctan\left(z\over U_{KK}\right)$
which is odd under $w\to -w$, and
\begin{eqnarray}
\alpha_\mu&=&\{\xi^{-1},\partial_\mu\xi \}={2i\over f_\pi}\partial_\mu \pi+\cdots\:,\nonumber\\
\beta_\mu&=& {1\over 2}[\xi^{-1},\partial_\mu\xi]={1\over
2f_\pi^2}[\pi,\partial_\mu \pi]+\cdots\:.
\end{eqnarray}
We recall from the previous eigenmode analysis by SS that
$\psi_{(2k+1)}(w)$ is even, while $\psi_{(2k)}(w)$ is odd under
$w\to-w$, corresponding to vector and axial-vector mesons
respectively.

Inserting this expansion into the action (\ref{5dfermion1}), and
using the properties $f_L(w)=\pm f_R(-w)$ as well as the properties
of $\hat\Psi_0$ and $\psi_{(n)}$ under $w\to-w$,
we obtain a 4D nucleon action
\begin{equation}
{\cal L}_4 = -i\bar B \gamma^\mu\partial_\mu B-im_B\bar BB+
{\cal L}_{\rm vector} +{\cal L}_{\rm axial}\:,
\end{equation}
with the four-dimensional nucleon mass $m_B$. This nucleon
mass will generally differ from the five-dimensional mass,
due to spread of the wavefunction $f_{L,R}$ along the fifth
direction. However, this difference arises only as a subleading
correction in large $N_c$ and large $\lambda$.
Writing out the interaction terms
explicitly, we have
\begin{equation}
{\cal L}_{\rm vector}=-i\bar B \gamma^\mu \beta_\mu
 B-\sum_{k\ge 0}g_{V}^{(k)} \bar B \gamma^\mu  a_\mu^{(2k+1)}
 B\:,\label{vector-coupling}
\end{equation}
and the nucleon couplings to axial mesons, including pions, as
\begin{equation}
{\cal L}_{\rm axial}=-\frac{i g_A}{2}\bar B  \gamma^\mu\gamma^5 \alpha_\mu B
-\sum_{k\ge 1} g_A^{(k)} \bar B \gamma^\mu\gamma^5 a_\mu^{(2k)} B\:,
\end{equation}
where various couplings constants $g_{V,A}^{(k)}$ as well as the
pion-nucleon axial coupling $g_A$ are calculated by suitable
wave-function overlap integrals. In the above expression, the
meson fields should be understood as being written in the nucleon
isospin representation.

The nucleon-meson interaction terms arise from two sources, namely
the magnetic-type direct coupling to the 5D gauge field strength
and the more conventional minimal coupling in the kinetic term.
The former comes with a coefficient $\rho^2_{baryon}/e^2$ in five
dimensions, which scales linearly with $N_c$.

The  minimal coupling contributions are summarized as
\begin{eqnarray}
g_{V,min}^{(k)}&=&\int_{-w_{max}}^{w_{max}} dw\,\left|f_L(w)\right|^2
\psi_{(2k+1)}(w)\:,\nonumber\\
g_{A,min}^{(k)}&=&\int_{-w_{max}}^{w_{max}} dw\,\left|f_L(w)\right|^2
\psi_{(2k)}(w)\:,\nonumber\\
g_{A,min}&=&2\int_{-w_{max}}^{w_{max}} dw\,\left|f_L(w)\right|^2
\psi_0(w)\:.
\end{eqnarray}
Since in the large $\lambda N_c$-limit, the nucleon wave-function
$f_L(w)$ tends to be symmetric under $w\to -w$, we see that
$g_{A}$ and $g_{A}^{(k)}$ receive small contributions from the
minimal 5D gauge interaction, in the large $\lambda N_c$ limit. On
the contrary, due to the even nature of $\psi_{(2k+1)}$, the
vector couplings $g_V^{(k)}$ receive an order one contribution
from the minimal interaction.

To isolate similar interaction terms from the 5D magnetic
coupling, we take the case of $(m,n)=(5,\mu)$, which becomes
\begin{equation}
-{\lambda
N_c(\rho_{baryon}M_{KK})^2\over 108\pi^3}\int d^4 x \int dw
\left[\left(2g_5(w)U(w)\over U_{KK} M_{KK}\right) \bar{\cal
B}\gamma^\mu\gamma^5(\partial_w A_\mu){\cal B}\right],
\end{equation}
where we have used
\begin{equation}
{1\over e^2(w)}=\frac{\lambda N_c}{108\pi^3}M_{KK}\frac{U(w)}{U_{KK}}\:.
\end{equation}
Defining the dimensionless number $C=\left({2\pi^2/3}\right){\lambda
N_c(\rho_{baryon}M_{KK})^2}/{ 108\pi^3}$,
we have contributions to $g_{V,A}^{(k)}$ and $g_A$ as follows,
\begin{eqnarray}
g_{V,mag}^{(k)}&=& 2C\int_{-w_{max}}^{w_{max}}
 dw \left(g_5(w)U(w)\over g_5(0)U_{KK}M_{KK}\right)\left|f_L(w)\right|^2\partial_w \psi_{(2k+1)}(w)\:,\nonumber\\
g_{A,mag}^{(k)}&=& 2C\int_{-w_{max}}^{w_{max}}
 dw \left(g_5(w)U(w)\over g_5(0)U_{KK}M_{KK}\right)\left|f_L(w)\right|^2\partial_w \psi_{(2k)}(w)\:,\nonumber\\
g_{A,mag}&=& 4C\int_{-w_{max}}^{w_{max}}
 dw \left(g_5(w)U(w)\over g_5(0)U_{KK}M_{KK}\right)\left|f_L(w)\right|^2\partial_w \psi_0(w)\:.
\end{eqnarray}
Note that the sizes of this integral behave oppositely when compared
to the similar overlap integrals for the minimal coupling term.
Again since in the large $\lambda N_c$-limit, the nucleon
wave-function $f_L(w)$ tends to be symmetric under $w\to -w$, and
since $\psi_{(2k+1)}$ ($\psi_{(2k)}$) is an even (odd) function of
$w$, the vector coupling contributions become relatively suppressed
in the large $\lambda N_c$-limit, while the axial couplings remain
order one times the large constant $C$. Using the estimate of
$\rho_{baryon}$ in Eq.~(\ref{size}), we find
\begin{equation}
C\simeq 0.18 N_c\: .
\end{equation}

With the present 5D effective theory approach, we have all mesons
encoded in a single $U(N_F=2)$ gauge field in five dimensions. In
particular, the iso-scalar mesons and iso-vector mesons arise from
a single $2\times 2$ gauge field. However, of these, only the
traceless part appears in the magnetic coupling since instanton
carries only non-Abelian field strength. Therefore, the iso-scalar
mesons and iso-vector mesons couple to nucleons differently. For the
iso-scalar mesons, such as for instance the $\omega$ meson in the
vector channel, only the minimal term contributes
\begin{eqnarray}
g_{A}^{(k)}\biggl\vert_{\rm iso-scalar}&=&g_{A,min}^{(k)}\biggl\vert_{\rm iso-scalar}\:,\nonumber\\
g_{V}^{(k)}\biggl\vert_{\rm iso-scalar}&=&g_{V,min}^{(k)}\biggl\vert_{\rm iso-scalar}\: .
\end{eqnarray}
But, for iso-vectors, we have contributions from both minimal and magnetic terms.
Thus, we must add
\begin{equation}
g_{A}=g_{A,mag}+g_{A,min}\:.
\end{equation}
and similarly for the vector and the axial vector in the iso-vector
\begin{eqnarray}
g_{A}^{(k)}\biggl\vert_{\rm iso-vector}&=&
\left[g_{A,mag}^{(k)}+g_{A,min}^{(k)}\right]\biggl\vert_{\rm iso-vector}\:,\nonumber\\
g_{V}^{(k)}\biggl\vert_{\rm iso-vector}&=&
\left[g_{V,mag}^{(k)}+g_{V,min}^{(k)}\right]\biggl\vert_{\rm iso-vector}\: .
\end{eqnarray}

Naively, since the coefficient for the magnetic term grows linearly
with $N_c$, one might be tempted to throw away the minimal coupling
contribution. However, the vector-like couplings and
axial-vector-like couplings behave quite differently. Note that the
part of the 5D magnetic term we employed above has an explicit
$\gamma^5$ while  the minimal term (in the gauge $A_5=0$) cannot,
yet both axial and vector-like coupling arise from each. It is only
because of the asymmetry between $f_L$ and $f_R$ that we can find
the axial interaction terms from the minimal coupling and the
vector-like terms from the magnetic coupling. This asymmetry is
strong when $\lambda N_c$ is small but diminishes as $\lambda
N_c\rightarrow \infty$. In other words, the wavefunction overlap
integral would be suppressed by $1/\lambda N_c$ for these
interactions with the ``wrong" number of $\gamma_5$.

For this reason, all axial couplings, $g_A$ and $g_A^{(k)}$ are
dominated, as expected, by the contribution from the magnetic terms,
whereas all vector-like couplings, $g_V^{(k)}$, will be dominated by
the contribution from the minimal couplings: at least in the large
$\lambda$ limit,
\begin{eqnarray}
g_{V}^{(k)}\biggl\vert_{\rm iso-vector}\simeq
g_{V,min}^{(k)}\biggl\vert_{\rm iso-vector}\: ;\quad
g_{A}^{(k)}\biggl\vert_{\rm iso-vector}\simeq
g_{A,mag}^{(k)}\biggl\vert_{\rm iso-vector},\: g_{A}\simeq
g_{A,mag}.
\end{eqnarray}

Finally let us note that we have neglected part of the magnetic
coupling in our discussion of the four-dimensional effective action,
namely those with two 4D indices on the Dirac matrices,
\begin{equation}
\bar{\cal B}\gamma^{\mu\nu}F_{\mu\nu}{\cal B} \:.
\end{equation}
When we reduce this dimensionally, we will find more couplings between
the nucleons and the infinite tower of mesons but with one more derivative
than the above Yukawa terms. Although they are higher power in usual
power counting, the suppressing mass scale would be at
most $M_{KK}$, so we expect these couplings to be very relevant
to physical processes which are measured up to several GeV. We
hope to come back to this aspect of holographic QCD in a later work.

\subsection{Vector Couplings: Iso-Scalar vs. Iso-Vector }

As we mentioned earlier, the couplings between massive vectors $a_\mu^{(2k+1)}$
and nucleons arise primarily from the minimal coupling in the large $\lambda$
limit. The leading coupling is then,
\begin{equation}
-\sum_{k\ge 0}g_{V}^{(k)} \bar B \gamma^\mu  a_\mu^{(2k+1)} B\:,
\end{equation}
where
\begin{equation}
g_V^{(k)} = g_{V,min}^{(k)}=\int_{-w_{max}}^{w_{max}} dw\,\left|f_L(w)\right|^2
\psi_{(2k+1)}(w)
\end{equation}
for $a_\mu^{(2k+1)}$ in the iso-scalar, while
\begin{equation}
g_V^{(k)} \simeq  g_{V,min}^{(k)}
\end{equation}
for $a_\mu^{(2k+1)}$ in the iso-vector in the large $\lambda$ approximation.

Since the iso-scalar and iso-vector couplings here have
the same origin in the five-dimensional dynamics, this immediately
implies a simple algebraic relations between the two classes of
couplings. Let us decompose the massive vectors as
\begin{equation}
a_\mu^{(2k+1)}=\left(\begin{array}{cc}1/2 &0 \\ 0&1/2\end{array}\right)
\omega^{(k)}_\mu+\rho_\mu^{(k)}
\end{equation}
into the trace part and the rest, where we wrote the gauge field in the
fundamental representation. This is how individual massless quark doublet
would see the vector mesons. However the baryon is made out of $N_c$ product
quark doublets, and we are considering the case of the doublet as the
smallest irreducible representation under $SU(N_F=2)$.
In the process, while the $SU(2)$ representation is kept small as such, the
trace part of the charge are simply added so
that the above decomposition actually appears for nucleons as
\begin{equation}
a_\mu^{(2k+1)}=\left(\begin{array}{cc}N_c/2 &0 \\ 0&N_c/2\end{array}\right)
\omega^{(k)}_\mu+\rho_\mu^{(k)}\:.
\end{equation}
We have been using the normalization of $SU(2)$ generators
consistently as $\tr \;T^aT^b=\delta^{ab}/2$, so the eigenvalues
for doublets are $\pm 1/2$.
Therefore, between the iso-scalar and the iso-vector, there is an
overall factor of $N_c$. In other words, we have the universal relation,
again in the large $\lambda$ limit
\begin{equation}
|g_{\omega^{(k)}NN}|\simeq  N_c \times
|g_{\rho^{(k)}NN}|\label{prediction-gomega}
\end{equation}
between the Yukawa couplings involving iso-scalar and iso-vector
vector mesons. Here $g_{vNN}$ denotes the Yukawa coupling between
the nucleon vector current and the canonically normalized vector
field $v$. Note that the relation (\ref{prediction-gomega}) is the
same as what one obtains in CQM. We will see how  the relation
(\ref{prediction-gomega}) fares with nature in Section
\ref{numerics} below.
%

\subsection{Pseudo-Vector Couplings}

An important observation to keep in mind here is that the normalization
condition of the eigenmode $\psi_{(n)}$ for $n\ge 1$ contains a factor of
$f_\pi$, so that of all quantities above, only $g_{A,mag}$ grows linearly
with $N_c$. Despite large $C$ value for large $N_c$, all other $g$'s are
order $(N_c)^0$ at most, and in fact suppressed  further by $1/\lambda$.
Nevertheless, it remains true that the contribution from the magnetic
coupling is dominant whenever present. Thus, depending on whether the
pseudo-vector is in the iso-scalar or in the iso-vector, we have the
following coupling
\begin{equation}
-\sum_{k\ge 1} g_A^{(k)} \bar B \gamma^\mu\gamma^5 a_\mu^{(2k)} B \:,
\end{equation}
where
\begin{equation}
g_A^{(k)}= g_{A,min}^{(k)}=\int_{-w_{max}}^{w_{max}} dw\,\left|f_L(w)\right|^2
\psi_{(2k)}(w)
\end{equation}
for iso-scalar part of the pseudo-vector $a_\mu^{(2k)}$ while
\begin{equation}
g_A^{(k)}\simeq  g_{A,mag}^{(k)}= 2C\int_{-w_{max}}^{w_{max}}
 dw \left(U(w)\over U_{KK}M_{KK}\right)\left|f_L(w)\right|^2\partial_w \psi_{(2k)}(w)
\end{equation}
for iso-vector part of the pseudo-vector $a_\mu^{(2k)}$.

\subsection{Axial Coupling to Pions and an $O(1)$ Correction \label{gA}}

\subsubsection{The leading ${ O}(N_C)$ term}\label{O(1)}
For  $g_A$, the leading contribution is $g_{A,mag}$, for which the
corresponding integral can be done exactly by using the explicit
form of $\psi_0(w)$ and also by approximating $g_5(w)\simeq
g_5(0)$. The latter approximation is harmless if $\lambda N_c$ is
sufficiently large. Since
\begin{equation}
\left(U(w)\over U_{KK}M_{KK}\right)\partial_w \psi_0(w)={1\over\pi} \:,
\end{equation}
we have
\begin{equation}
g_{A,mag}={4C\over\pi}\simeq
0.7\times\frac{N_c}{3} \:.\label{leading-result}
\end{equation}
While this depends on the substitution of $g_5(w)\rightarrow g_5(0)$,
the result is robust as long as $f_L(w)$ is sufficiently
localized at $w=0$. In turn, this  is guaranteed by arbitrarily large
$\lambda N_c$.

The subleading contribution, $g_{A,min}$ is at most order $1/\lambda
N_c$, and thus is negligible in the present AdS/CFT limit.
\subsubsection{The ${ O}(1)$ correction}\label{shift}
As mentioned in Section \ref{5DB}, the collective quantization that
led to (\ref{leading-result}) was based on the mathematical
manipulation of ANW that consisted of performing the isospin
rotation $A=a_4+ia_i\tau_i$ (with $\sum_i a_i^2=1)$ of the soliton
and evaluating the corresponding element of the orthogonal space
rotation group given by
 \be
R_{ij}=\frac 12{\Tr}{\tau_i A\tau_j A^\dagger}.\label{collective-q}
 \ee
We can exploit the equivalence of the constituent quark model (CQM)
and the Skyrmion in the large $N_c$ limit~\cite{manohar-lecture} to
obtain an $1/N_C$ correction to the leading term while fermion loops
are kept suppressed. Briefly, the reasoning goes as follows.
 \begin{enumerate}
 \item
We first note that the collective quantization of the instanton we
are dealing with involves, among various collective coordinates, the
same isospin rotation (\ref{collective-q}) as in the Skyrme model.
This can be seen in the collective quantization of the instanton by
Hata et al.~\cite{Hata:2007mb}. Now the ANW quantization is known to
give the ${\calO} (N_c)$ term to $g_A$ which is identical to what is
obtained in the large $N_c$ limit of CQM~\cite{manohar-lecture}.
 \item
A general large-$N_c$ QCD analysis shows that $g_A$ has the large
$N_c$ expansion~\cite{dashen-manohar-gA}
 \be
g_A=\alpha\left(\frac{N_c+\beta}{3}\right)+\gamma\frac{1}{N_c}+\cdots\label{DM-theorem}
 \ee
where $\alpha$, $\beta$ and $\gamma$ are constants independent of
$N_c$ and the ellipsis stands for higher $1/N_c$ terms. An
important point to note here is that fermion (quark) loop
corrections first appear at ${\calO}(1/N_c)$ and not at
${\calO}(1)$. This means that the constant $\beta$ survives
``quenching," that is, it has {\it no dynamical loop effects}.
 \item
While general considerations leave the coefficient $\beta$
undetermined, the CQM, however, gives a simple result coming from a
simple (group-theoretic) book-keeping,
 \be
\beta=2.\label{b}
 \ee
One might a priori think that the Skyrmion model needs not give
the same value. However it has been shown by a detailed group
structure of the spin-isospin operator involved in the Skyrmion --
and likewise in the instanton baryon -- that the result (\ref{b})
$does$ hold~\cite{amadoetal}.~\footnote{Briefly the argument is as
follows~\cite{amadoetal}. The spin-isospin structure of the
hedgehog ansatz adopted for the instanton (Skyrmion) suggests that
the soliton is a $U(4)$ coherent state in the large $N_c$ limit.
By realizing the soliton algebra in terms of $N$ ``interacting
bosons" (with $N=N_c$) familiar in nuclear and molecular physics,
projecting out the good spin and isospin of the nucleon in the
matrix element of the axial-current operator is made both direct
and simple. This permits us to calculate the leading $1/N_c$
correction based solely on symmetry consideration without
involving any dynamical calculations.} Exactly the same argument
holds for the iso-vector dipole magnetic moment operator and will
be applied later in the next section.
\end{enumerate}

If one shifts $N_c$ to $N_c+2$ as argued above, we can include the
${\calO}(1)$ correction to (\ref{leading-result}) and obtain
 \be
g_A\approx 0.7\left(\frac{N_c+2}{3}\right)\approx
1.17\:,\label{shifted-result}
 \ee
which is expected to be reliable up to ${\cal O}(1/N_c^2)\approx
10\%$. An interesting observation to make at this point is that
the instanton baryon predicts $\alpha\approx 0.7$ in
(\ref{DM-theorem}) which is close to the chiral perturbation
theory prediction $\alpha_{\chi PT}\approx 0.75$~\cite{gA-chpt}.
Another observation is that the ``probe approximation" involved in
the SS model appears to be equivalent to the quenched
approximation in lattice calculations. The quenched lattice
calculation contains no fermion loops while containing all orders
of $\lambda$ and $1/N_c$ pertaining to gluons. We conjecture that
the quenched lattice result differs from the instanton result
(\ref{shifted-result}) only at the next order, i.e.,
${\calO}(1/N_c^2)$ relative to the leading order. This conjecture
is numerically supported in that the quenched lattice
result~\cite{dongetal,ohtaetal} is quite close to
(\ref{shifted-result}), and furthermore the unquenched
calculation~\cite{ohtaetal} with dynamical quarks agrees closely
with the quenched result indicating that the higher order $1/N_c$
corrections are not big.

\section{Numerical Estimates and Extrapolations \label{fat}}

\subsection{Numerics and Subleading Corrections}\label{numerics}

In this effective theory approach,  we consider the
five-dimensional baryon as point-like, which is justified by the
large 't Hooft coupling $\lambda=\g_{YM}^2N_c$. However, if we
wish to extrapolate the result to finite $\lambda \sim 17 $
regime, we cannot neglect the size of the instanton. Thus, we
cannot say that the above can be extrapolated to a realistic QCD
regime with justification. We will come back to the size issue in
the last part of this section. However, with this caveat in mind,
we wish to extrapolate the effective theory to the realistic
regime and try to see how leading corrections would behave
qualitatively. Typical quantities we must know to compare with
nature are the trilinear couplings, namely $g^{(k)}_{V,min}$,
$g^{(k)}_{V,mag}$, $g^{(k)}_{A,min}$, $g^{(k)}_{A,mag}$, as well as
$g_{A,mag}$  and $g_{A,min}$. In the previous section,
we outlined the large $N_c$ and large $\lambda$ behavior of these
couplings which must be corrected as we approach realistic
regimes.

\begin{table}[h]
\begin{tabular}{|c|c|c|c|c|c|c|c|}
\hline
$\lambda N_c $& ${m_B / M_{KK}} $& $ {m_B / f_{\pi}}$
& $g_{A,min}$ &$g_{V,min}^{(0)}$ & $g_{V,min}^{(1)}$ & $g_{V,mag}^{(0)} (N_c=3)$ & $g_{V,mag}^{(1)} (N_c=3)$ \\
 \hline\hline
10& 1.37 & 22.2 & 0.171 & 11.6 & 3.87 & -1.81& -2.76\\
 \hline
20 & 1.52 &17.5 & 0.161 & 8.70 & 3.61 & -1.34  & -2.55\\
\hline 30 & 1.66 & 15.6 & 0.152 & 7.35 & 3.45 & -1.10  & -2.35\\
\hline 40 & 1.80 &14.6 & 0.143 & 6.52 & 3.32 &-0.93  & -2.16 \\
\hline 50 & 1.93 &14.0 & 0.135 & 5.93 & 3.22 &-0.82  & -1.98 \\
\hline 60 &2.06 & 13.7 &0.129 & 5.49 & 3.13  & -0.72  & -1.87\\
\hline 80 & 2.32 & 13.3 & 0.117 & 4.85 & 2.98 & -0.59 & -1.63\\
\hline 120 & 2.82 & 13.2 & 0.099 & 4.07  & 2.74 & -0.42 & -1.27 \\
\hline 160 & 3.30 & 13.4 & 0.086 & 3.59  & 2.56 & -0.33 & -1.03\\
\hline 200 & 3.79 & 13.7 & 0.076 & 3.24  & 2.40 & -0.27 & -0.86 \\
 \hline
\end{tabular}
\caption{\small Numerical result for  $g_{A,min}$, the axial
pion-nucleon-nucleon coupling, and the couplings to the lowest two
vector mesons. Here we used $C\simeq 0.18 N_c$ with $N_c=3$ for
the evaluation in the last two column. The realistic regime should
be chosen so that ${f_\pi / M_{KK}}=\sqrt{\lambda N_c / 54\pi^4}$
fits with experimental values for these two scales. The resulting
$\lambda N_c$ lies somewhere around 50. For $g_{V,mag}$'s, we
approximated $g_5(w)/e(w)^2=g_5(0)/e(0)^2$ which may not be
justifiable in the present range of $\lambda N_c$ values. }
\label{table1}
\end{table}


The main object we need to understand in order to compute these couplings
is the wavefunction of the nucleon $f_{L,R}(w)$. For efficient numerical
estimates, we scale out dimensionful parameters from the spinor equations
by introducing
dimensionless variables $\tilde w=M_{KK}w$, $\tilde U={U/
U_{KK}}$, and $\tilde z={z/ U_{KK}}$. These are related as
\begin{equation}
\tilde w = \int_0^{\tilde z}\, {d\tilde z \over \left[1+\tilde
z^2\right]^{2\over 3}} = {3\over 2}\int_1^{\tilde U}\,{d\tilde
U\over \sqrt{\tilde U^3 -1}}\:.
\end{equation}
In terms of these variables,
we  have
\begin{equation}
m_b(z)=m_B^{(0)}\cdot\tilde U+m_0^e=M_{KK}\cdot\left({\lambda N_c\over 27\pi}\tilde U(\tilde
w)+\epsilon  N_c\right)
\end{equation}
with $\epsilon\equiv \sqrt{2/15}\simeq 0.37$.
After dividing the eigenvalue
equation (\ref{eigeneq}) by $M_{KK}^2$, we arrive at
\begin{equation}
\left[-\partial^2_{\tilde w} -{\lambda N_c\over 27\pi}\,
\partial_{\tilde w}\tilde U({\tilde w})+\left( {\lambda N_c\over 27\pi}\,\tilde U(\tilde w)+\epsilon N_c
\right)^2\right] f_L(\tilde w)=\left(m_B\over M_{KK}\right)^2
f_L(\tilde w)\:,\label{redeigeneq}
\end{equation}
and the wave-function $f_L(\tilde w)$ does not depend on the scales.
Since $\psi_0(w)$
is also a universal function in terms of
our dimensionless variables, the two axial coupling contributions,
$g_{A,min}$ and $g_{A,mag}$, are indeed
functions of $\lambda N_c$ and $N_c$. Specifically, the previous
formula tells us that $g_{A,min}$ is a function of $\lambda N_c$
and $g_{A,mag}$ depends on $N_c$ only.

\begin{figure}[t]
\begin{center}
\scalebox{1.12}[1.2]{\includegraphics{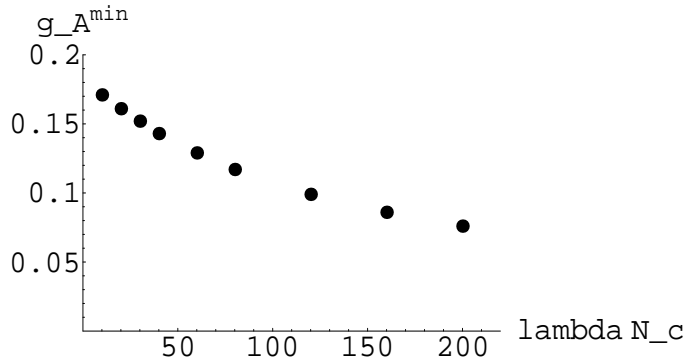}}
\par
\vskip-2.0cm{}
\end{center}
\caption{\small Plot of $g_{A,min}$ versus $\lambda N_c$. } \label{fig}
\end{figure}

We solve  $f_{L}(\tilde w)$ and its eigenvalue ${m_B/ M_{KK}}$
numerically for a given value of $\lambda N_c$ using shooting
method. As mentioned before, the Coulomb energy part, $C_0 N_c$, is
subleading and negligible in the 't Hooft limit. For large $\lambda
N_c$, the effective potential in (\ref{redeigeneq}) is very steep
and the wave-function would tend to localize at the minimum point
which scales as ${\tilde w}_{min}\sim {\cal O}((\lambda N_c)^{-1})$.

For instance, we can see that $g_{A,min}$ (and $g^{(k)}_{A,min}$)
are proportional to the asymmetry of $|f_{L}(\tilde w)|^2$ in
$\tilde w$ for small ${\tilde w}_{min}$, we conclude that
$g_{A,min}$ also scales as ${\cal O}((\lambda N_c)^{-1})$ for large
$\lambda N_c$. Our numerical result is shown in Table 1, and the
values of $g_{A,min}$ for large $\lambda N_c$ confirm this
expectation. The same is true of $g^{(k)}_{V,mag}$, relative to
$g^{(k)}_{V,min}$. Table \ref{table1} provides some numerical values
for $g_{A,min}$, $g^{(k)}_{V,min}$, and $g^{(k)}_{V,mag}$. The first
represents a subleading correction to the axial coupling between
pions and nucleons, whereas $g^{(k)}_V$ are quantities which are
also well-measured via scattering processes of nucleons.

Before proceeding further, however, we must warn the readers of
another approximation we took which goes beyond the usual large
$N_c$ and large $\lambda$ limit. Note that our computation in
section 4 revealed the value of coupling $g_5$ at $w=0$. Extending
this to a bona-fide function of $g_5(w)$ has so far proven very
difficult. While some quantities, such as $g_{A,mag}$, is
insensitive to the detailed form of this function, generic numerical
estimate requires its precise form. Roughly speaking, this problem
will become more and more severe for large values of $k$ since its
wavefunction would be spread more and more away from the origin
$w=0$. Also the smaller the value of $\lambda N_c$, the less
reliable will be our estimate since $f_{L,R}(w)$ will be also spread
more and more away from the origin. This is a technical problem that
affects all terms arising from the magnetic terms. For numerical
estimates of $g^{(k)}_{V,mag}$ here and later in section 7, we chose
to sidestep the issue by replacing $g_5(w)/e(w)^2$ by its value at
the origin $g_5(0)/e(0)^2$.

Using Table 1, in conjunction with the results of previous section
on leading large $N_c$ behaviors, we can make semi-quantitative
estimates of $g_A$ and $g_{V NN}$ for $V=\rho, \omega$ and compare
with nature. To do this, we adopt the parameters $M_{KK}$ and
$\lambda N_c$ fixed by the pion decay constant $f_\pi\approx 86
\sim 93$ MeV~\footnote{$f_\pi$ is 86 MeV for $m_\pi=0$ and 93 MeV
for $m_\pi\approx 140$ MeV.} and the $\rho$-meson mass in the
meson sector~\cite{sakai-sugimoto}, i.e., $M_{KK}\approx .94$ GeV
and $\lambda N_c=50$.
 \begin{itemize}
\item {\bf The axial coupling constant}:
\vskip 0.1cm Adding the subleading contribution $g_{A,min}$ of
Table \ref{table1} to the leading term (\ref{shifted-result}), we
have for $\lambda N_c=50$,
 \be
 g_A\approx  1.30-1.31
 \ee
which compares well with the experimental value $g_A^{exp}=1.2670\pm
0.0035$. As discussed in Section \ref{O(1)}, there are indications
from lattice calculations that higher-order $1/N_c$ corrections or
``unquenching" are suppressed. The same suppression seems to be
taking place in our calculation.
\item {\bf The ${\rho NN}$ and ${\omega NN}$ coupling constants:}
\vskip 0.1cm Consider the lowest members of the tower $V=\rho,
\omega$ that correspond to $k=0$ in Eq.(\ref{vector-coupling}).
The leading order relation (\ref{prediction-gomega}) will be
spoiled at the subleading order since the magnetic term
contributes only to the $\rho NN$ coupling. From Table
\ref{table1}, we have for $\lambda N_c=50$
 \bear
g_{\rho NN} &\approx& 3.6 \nonumber\\
g_{\omega NN} &\approx& 12.6\label{theory}
 \eear
Thus the relation (\ref{prediction-gomega}) is modified to
 \be
{\cal R}\equiv \frac{g_{\omega NN}}{3g_{\rho NN}}\approx
1.2\label{ratio}
 \ee
roughly independently of $\lambda N_c$.  We should stress that while
the sign of $g_{V,mag}$ is robust, the approximation that goes into
the estimate of $g_{V,mag}$ is uncertain, so we cannot take the
numerical values too seriously. {\it However considering that there
are no theoretical estimates -- instead of fits to experiments -- of
the above quantities, we offer (\ref{theory}) and (\ref{ratio}) as
the first theoretical prediction of those quantities.}

There are no direct experimental determinations of these constants.
However indirect ``empirical" values have been extracted from
various sources including precision fits to nucleon-nucleon
scattering phase shifts up to lab energy $\sim 350$ MeV using
one-boson-exchange potentials. In addition, purely phenomenological
potentials parameterized with a large number of parameters fit to
phase shifts can be translated into the form of boson-exchange
potentials and provide information on the effective
constants~\cite{riska}. Unfortunately since no direct determination
from experimental data are feasible, the numbers extracted from such
analysis are far from unique and in fact they can vary quite
widely~\footnote{Over the three decades from the early efforts in
1970's~\cite{hoehler-pietarinen} through extensive studies in 1980's
and 1990's~\cite{nijmegen} to the most recent
ones~\cite{machleidt,gross07}, there seems to be little convergence
on both $g_{\rho NN}$ and $g_{\omega NN}$ except that $g_{\omega
NN}> 3g_{\rho NN}$.}. With this caveat in mind, let us quote the
ranges of values found in the literature. They are
 \bear
g_{\rho NN}^{emp}\approx  4.2-6.5, \ \  {\cal R}\approx 1.1-1.5.
\label{empirics}
 \eear

Although the individual values for $g_{V NN}$ extracted empirically,
e.g., (\ref{empirics}), are subject to uncertainties mentioned
above, it has been a mystery why NN phase shifts $invariably$
required that $g_{\omega NN}$ be larger than the CQM prediction $3
g_{\rho NN}$. Remarkably, this observation is naturally explained in
the holographic QCD model as one can see in (\ref{ratio}) although
the quantitative comparison may not be meaningful as mentioned
above. One should also note that (\ref{theory}) violates what is
referred to as ``universality," namely, $g_{\rho \pi\pi}=g_{\rho
NN}$, as empirically $g_{\rho\pi\pi}\approx 6$ which is closer to
$g_{\rho,min}$ in Table \ref{table1} for the relevant range of
$\lambda N_c$. The source for this violation is in the magnetic
contribution $g_{\rho, mag}$ which is also responsible for the ratio
${\cal R}$ to deviate from 1.

 \end{itemize}
\subsection{An Issue with Extrapolation: Size of the Baryon }

So far, we studied static and dynamical behaviors of baryons by
starting with small instantons with fundamental string hairs, in
the very large 't Hooft coupling limit. However, for intermediate
values of 't Hooft coupling,  the
story has to change qualitatively. Recall that the size of the
instanton
\begin{equation}
\frac{9.6}{M_{KK}\sqrt{\lambda}}
\end{equation}
can be fairly large for the 't Hooft coupling of order 10.
As we consider larger and larger instanton size, however, the computation
leading to this estimate loses the validity. In particular,
the instanton energy from Yang-Mills action is affected drastically.
The effective mass from the instanton density scales with $1/e^2(w)
\propto (U/U_{KK})$. While we used the leading behavior
$U/U_{KK}\sim 1+\frac13 M_{KK}^2w^2$ for small $w$, $1/e^2(w)$ is
in fact  divergent as $w\rightarrow\pm w_{max}\simeq 3.64/M_{KK}$.
With $\lambda\sim 17$, the diameter of the instanton according to the
above estimate is about $2/M_{KK}$, which immediately shows that we are well
out of region of validity. The extra energy  in Eq.~(\ref{mass}) is
a gross underestimate.

Also the Coulomb energy  Eq.~(\ref{C}) can be seen to be modified.
It treats the five-dimensional gauge field as a massless gauge
field living in flat $R^{4+1}$. In reality, for configurations of
size comparable to $1/M_{KK}$, this is not the right picture.  In
particular, the increasing value of $1/e(w)^2$ outward along $w$
effectively makes the physics four-dimensional, where the
five-dimensional vector field should be replaced by an infinite
tower of massive vector mesons. The lightest has the mass $\sim
0.8 M_{KK}$, so the Coulomb energy estimated in
Section.~\ref{small} must be augmented by an exponential
suppression as well, changing to the power-exponent,
\begin{equation}
\sim \frac{M_{KK}e(0)^2 N_c^2}{ \rho}e^{-0.8\rho M_{KK}} \:.
\end{equation}
Thus, Eq.~(\ref{C}) is a bit of overestimate for large sizes.

Making these estimates more precise requires further effort. The
main difficulty comes from the fact that we cannot use the usual
self-dual instanton on $R^4$ to estimate the potential which is to
be minimized. The  problem is that the latter does not satisfy
physical boundary condition at $w=\pm w_{\max}$ and that, even if we
wish to use the usual instanton only as an approximate trial
configuration, the divergent  $1/e^2(w)$ at the boundary makes the
energy of such configuration always infinite. What we need is a
reasonable trial configuration whose gauge field strength vanishes
very fast as $w\rightarrow\pm w_{max}$. These difficulties were in
fact also present in the estimate of Section.~\ref{small} as well,
which we ignored without justification, but it is unlikely that this
detail would change the large 't Hooft coupling behaviors since the
instanton involved is very small. Here we need to correct it since
we are now talking about instantons whose size is comparable to the
length scale of the fifth direction and since the order one factors
are more important.

At the end of the day, however,
the combined effect has to be that the instanton gets stabilized
at much smaller size than predicted by the naive extrapolation
of the size estimate we used. We anticipate that the size would
be stabilized to be no larger than $1/M_{KK}$. Once we are in
this regime, on the other hand, the strategy we followed loses
all of its validity, and it would be misleading to proceed in
the same manner, only with the size of the instanton modified.
As long as we are interested in interactions of the baryons
with other fields in this theory, we propose that the right thing
to do is to set up the effective field theory in the large $\lambda$
limit, where all computations we carried out are well-justified, and
extrapolate only at the end of the day when comparing scattering
amplitudes. When we consider four-dimensional processes which
are not very sensitive to the  `t Hooft coupling in the large $N_c$
expansion, this strategy is most likely to be successful. We believe
this is the reason why our approach produced reasonable numbers
even when compared to experimental values.

\section{Electromagnetic Interaction and Vector Dominance}\label{VD}

A prominent feature of the holographic dual QCD is that its interaction
with electromagnetic field is vector dominated.
Let us first consider the situation with pions in the SS
holographic model of QCD. There, the electromagnetic form factor of the pion is
given by the entire tower of the vector mesons~\cite{sakai-sugimoto},
 \be
F_1^\pi (q^2)=\sum_{k=0}^\infty
\frac{g_{v^{(k)}}g_{v^{(k)}\pi\pi}}{q^2+m_{v^{(k)}}^2} \:.
 \ee
The quantities  $g_{v^{(k)}\pi\pi}$
are the trilinear couplings between pions and the vector mesons. The
vector meson $v^{(k)}$ are defined as linear combinations of $a^{(2k+1)}$
and ${\cal V}$, as will be shown explicitly below. Accordingly its mass
$m_{v^{(k)}}$ is $m_{2k+1}$ in our notation. The parameters $\zeta_k\equiv
g_{v^{(k)}}/m_{2k+1}^2$, which will be introduced shortly, encode how
the photon field mixes with the massive vector mesons.

This form factor shows that there is no direct contact charge and arises because
all electromagnetic interaction of pions necessarily goes through
intermediate vector mesons.
The charge form factor evaluated at $p^2=0$ is the charge of the
particle, and thus we must have the normalization
\be\label{sumrule}
F_1^\pi (0)=\sum_{k=0}^\infty
\frac{g_{v^{(k)}}g_{v^{(k)}\pi\pi}}{m_{v^{(k)}}^2}=1 \:.
\ee
In the SS model of QCD, this sum rule is a mathematical consequence of
the completeness of the normalizable eigenmodes along the fifth direction.
This sum rule has been also checked numerically, and
for pions, it has been shown  that the sum rule
(\ref{sumrule}) is saturated within less than 1\% by the first four
low-lying vector mesons in the $\rho$ quantum number. However the
lowest member $\rho$ exceeds the sum rule by $\sim 30$\%, so the
next three are important in the sum rule.

In the following we will sketch how this vector dominance arises for
pions and how this generalizes $naturally$ to nucleons in our
current effective action approach. Several analog of the above sum
rule will also appear naturally from the completeness of eigenmodes,
and we will see that again truncation down to the first four massive
modes saturates these sum rules for nucleons within 1\%.

\subsection{The Vector Dominance for the Nucleons}\label{VD-N}

As we saw before,
vector mesons, $a^{(2k+1)}_\mu$, and axial vector mesons,
$a_\mu^{(2k)}$, arise as massive KK modes and exhaust all
normalizable eigenmodes of the vector field which can be used upon
dimensional reduction. Among the normalizable degrees of freedom,
there is no room for photon field. Instead, the coupling to the photon field
must be read out via the usual AdS/CFT prescription by
computing an appropriate current to be matched with an external $U(1)_{em}$ field,
${\cal V}$. The latter, in our language, shows up as
non-normalizable term added to the $i\beta_\mu(x)$ term\footnote{
One can also introduce an axial vector ${\cal A}_\mu(x)$ added to $i\alpha_\mu(x)$. This would
be relevant to the coupling of the hadrons to $SU(2)_{weak}$ but
here we will not consider it.}
\begin{equation}
A_\mu(x;w)= i\alpha_\mu(x)\psi_0(w)+{\cal V}_\mu(x)+i\beta_\mu(x) +\sum_n
a_\mu^{(n)}(x)\psi_{(n)}(w) \:.
\end{equation}
Upon integrating
over the fifth direction, this generates a term of the type
\begin{equation}
\int dx^4\;{\cal V}_\mu J^\mu
\end{equation}
giving us the vertex $J$. After specializing ${\cal V}$ to a $U(1)$
subgroup, appropriately chosen to be consistent with four dimensional
physics, electromagnetic vertices can be read out.

For a generalization to nucleons,  it is instructive to recall how
the vector dominance came about in the meson sector in the SS model.
If we keep both the vector mesons and
this external vector we have the following general structure
of the lagrangian,
\begin{equation}
-\sum_n {\rm
tr}\;\left[\frac12|da^{(n)}|^2+m_n^2|a^{(n)}|^2\right] -\sum_k
\zeta_k\;{\rm tr}\;\langle da^{(2k+1)},d({\cal V}+i\beta)\rangle
\end{equation}
with
\begin{equation}
\zeta_k=\int dw\frac{1}{2e(w)^2}\,\psi_{(2k+1)}(w) \:,
\end{equation}
which is related to $g_{v^{(k)}} $ of Sakai and Sugimoto as
\begin{equation}
g_{v^{(k)}}=m_{2k+1}^2\zeta_k \:.
\end{equation}
The reason why only the vectors and not the axial vectors shifts by
${\cal V}$ is clear from the form $\lambda$'s, for the eigenmodes for the
axial vectors are odd functions.

For canonical forms
of the kinetic terms, then we must introduce  shifted vector fields
\begin{equation}
v^{(k)}=a^{(2k+1)}+\zeta_k ({\cal V}+i\beta) \:,
\end{equation}
where we now have
\begin{equation}
\sum_k {\rm tr}\;\left[-\frac12|da^{(2k)}|^2-m_{2k}^2|a^{(2k)}|^2 -
|dv^{(k)}|^2 - m_{2k+1}^2|v^{(k)}-\zeta_k ({\cal
V}+i\beta)|^2\right]
\end{equation}
up to a term that additively renormalizes the kinetic term of ${\cal
V}$. This induces a quadratic vertex between vector mesons and the
external gauge field ${\cal V}$, which induces in turn indirect
couplings between ${\cal V}$ and pions. After some computation, one
can show for the SS model that the cubic couplings between pions and
these vectors are organized into  the form
\begin{equation}
g_{v^{(k)}\pi\pi}\,\tr
\left(v^{(k)}_\mu[\pi,\partial^\mu\pi]\right).
\end{equation}
These cubic couplings between pions and $v^{(k)}$ plus the
quadratic mixing between $v^{(k)}$ and ${\cal V}$ generates an effective
cubic interaction of pions with ${\cal V}$, and summing over all
intermediate vector mesons generates the form factor $F_1^\pi$. In
particular, the zero momentum limit of this form factor is the
electromagnetic charge of the pion, and the sum rules ensure
consistency such as charge quantization.

Now let us see how this mixing of vector fields enters the coupling
of baryons with electromagnetic vector field ${\cal V}$. Seemingly
this case is very different from that of pions. For one thing, we
have a minimal interaction term between nucleons and the 5D gauge
field, $A$, and this is inherited by ${\cal V}$ without
modification, since ${\cal V}$ is simply the non-normalizable part
of $A$. Thus it may seem that we have a point-like interaction
between baryons and ${\cal V}$, precluding the notion of vector
dominance in the form factor. However, nucleons also couple minimally
to the 4D massive vectors, $a^{(2k+1)}$, which mix with ${\cal V}$
in the propagator. They show up in the baryon effective Lagrangian
as
\begin{equation}
\int dw \;\bar {\cal B}\gamma^mA_m{\cal B}=
\bar B\gamma^\mu {\cal V}_\mu B+\sum_k g^{(k)}_{V,min} \bar B\gamma^\mu a^{(2k+1)}_\mu B+\cdots \:,
\end{equation}
where the ellipsis denotes axial couplings to axial vectors
as well as  coupling to pions via $\alpha_\mu$ and $\beta_\mu$.
There should be additional contribution from the 5D magnetic
coupling, shifting $g_{V,min}$, to which we will come back shortly.

Alternatively, we may use the canonically normalized vectors
$v^{(k)}$ instead, where we have the vector-current couplings
of type
\begin{equation}
\bar B\gamma^\mu {\cal V}_\mu B+\sum_k g^{(k)}_{V,min}
\bar B\gamma^\mu (v^{(k)}_\mu-\zeta_k {\cal V}_\mu ) B+\cdots \:.
\end{equation}
On the other hand,
\begin{eqnarray}
\sum_kg^{(k)}_{V,min}\zeta_k&=&\sum_k \int dw'\;|f_L(w')|^2\psi_{(2k+1)}(w')\times
\int dw\;\frac{1}{2e(w)^2}\,\psi_{(2k+1)}(w)\nonumber\\
&= &\sum_n \int dw'\;|f_L(w')|^2\psi_{(n)}(w')\times
\int dw\;\frac{1}{2e(w)^2}\,\psi_{(n)}(w)\nonumber\\
&=& \int dw'\;|f_L(w')|^2\times \int
dw\;\frac{1}{2e(w)^2}\,\sum_n\psi_{(n)}(w)\psi_{(n)}(w')\:,
\end{eqnarray}
where the second step makes use of the fact that
$1/e(w)^2$ is an even function. Using the completeness of
the  normalizable eigenmodes $\psi_{n}$, we find that
\begin{eqnarray}
\sum_k g^{(k)}_{V,min}\zeta_k = \int dw'\;|f_L(w')|^2\times
\int dw\;\delta(w-w')=\int dw'\;|f_L(w')|^2=1 \:,
\end{eqnarray}
which implies the crucial sum rule,
\begin{eqnarray}
\sum_k g^{(k)}_{V,min}\zeta_k =1 \:.\label{crucialSR}
\end{eqnarray}
Therefore, in this shifted basis, we have the charge form factor
\begin{equation}
\bar B\gamma^\mu {\cal V}_\mu B+\sum_k g^{(k)}_{V} \bar B\gamma^\mu (v^{(k)}_\mu
-\zeta_k{\cal V}_\mu) B+\cdots
= \sum_k g^{(k)}_{V} \bar B\gamma^\mu v^{(k)}_\mu B+\cdots
\end{equation}
As in the case of the pion, we can see that the cubic
electromagnetic interaction is mediated entirely by intermediate
massive vector mesons, rendering the nucleon form factors entirely
vector-dominated. This aspect will be highlighted in section 9.

Of course, this choice of basis is only for the sake of clarity.
The $\{{\cal V}; v^{(k)}\}$ basis is such that the mixing between
${\cal V}$ and massive vector meson is maximal in the zero momentum
limit, and thereby exhibits clearly how the minimal coupling to
photon field is replaced by the mediation via massive vector mesons.
However, the physics should be independent of such choices. In the
following, we will compute the charge form factor and the Pauli
form factor explicitly in the original $\{{\cal V}; a^{(2k+1)}\}$ basis
and see how the physical quantities bear out the notion of the vector dominance.

\subsection{Charge Form factor $F_1$}

To be more precise, let us compute the effective 3-point vertex of
type $\bar B\gamma^\mu {\cal V}_\mu B$. Let us first put a cut-off along
the fifth direction integrals, which effectively make ${\cal V}$
dynamical with a large kinetic term
\begin{equation}
\frac{L}{2}\,{\rm tr}\;|d{\cal V}|^2
\end{equation}
for some large number $L$, whose precise value will not matter.
The propagator for $\{{\cal V}; a^{(2k+1)}\}$ is such that
\begin{eqnarray}
\langle {\cal V}(q){\cal V}(-q)\rangle &\sim & \frac{i}{Lq^2
-\sum_k \zeta_k^2q^4/(q^2+m_{2k+1}^2)}\:,\nonumber \\
\langle a^{(2k+1)}(q) {\cal V}(-q)\rangle &\sim &-\langle {\cal V}(q){\cal V}(-q)\rangle \times
\frac{\zeta_k q^2}{q^2+m_{2k+1}^2}\:.
\end{eqnarray}
The electromagnetic form factor $F_1$ can be found from this by
computing tree-level correlator, $\langle \bar B\gamma^\mu {\cal V}_\mu B\rangle$,
and amputating the external lines. The resulting charge form factor $F_{1,min}$
which arises from the minimal interaction term is
\begin{equation}
F_{1,min}(q^2)=1-\sum_k \frac{g^{(k)}_{V,min}\zeta_k
q^2}{q^2+m_{2k+1}^2}= \sum_k \frac{g^{(k)}_{V,min}\zeta_k m_{2k+1}^2
}{q^2+m_{2k+1}^2}= \sum_k \frac{g_{v^{(k)}}
g^{(k)}_{V,min}}{q^2+m_{2k+1}^2}\label{chargeff}
\end{equation}
up to the electromagnetic charge operator.
We used the sum rule $\sum_k g^{(k)}_{V,min} \zeta_k=1$ and the definition
$g_{v^{(k)}}=\zeta_k m_{2k+1}^2$. The first expression is natural in the
$\{{\cal V};a^{(2k+1)}\}$ basis while the second expression is natural in the
$\{{\cal V};v^{(k)}\}$ basis. The result is, of course, independent of the basis
choice.

Note that there is no contact charge in the baryon, which would have
resulted in $F_1(\infty)\neq 0$. However, since the holographic
model used is defined by the mass scale $M_{KK}\sim 1$ GeV, our form
factor does not have the correct asymptotic behavior of perturbative
QCD, $F_1 (q^2)\sim 1/q^4$~\cite{lepage-brotsky}. This must be
implemented by hand if one wanted to fit the experimental data at
large momentum transfers.

The actual charge form factor picks up an additional contribution
from the magnetic coupling, since the latter contributes couplings
$g^{(k)}_{V,mag}$ between nucleon current and massive vector mesons
as well. This does not induce an additional electric charge (as it
should not) given the charge quantization, and this happens as a
consequence of another sum rule:
\begin{eqnarray}
&&\sum_kg^{(k)}_{V,mag}\zeta_k\nonumber \\
&
=&\sum_k \int dw'\;\left(g_5(w')U(w') \over g_5(0)U_{KK}M_{KK}\right)|f_L(w')|^2\partial_{w'}\psi_{(2k+1)}(w')\times
\int dw\;\frac{1}{2e(w)^2}\,\psi_{(2k+1)}(w)\nonumber\\
&=&\int dw'\;\left(g_5(w')U(w') \over g_5(0)U_{KK}M_{KK}\right)|f_L(w')|^2\times
\int dw\;\partial_{w'}\delta(w-w')\nonumber\\
&=&-\int dw'\;\partial_{w'}\left[\left(g_5(w')U(w') \over g_5(0)U_{KK}M_{KK}\right)|f_L(w')|^2\right]
=0 \:.
\end{eqnarray}
The contribution to the charge form factor from the magnetic coupling is
then,
\begin{equation}
F_{1,mag}(q^2)=-\sum_k \frac{g^{(k)}_{V,mag}\zeta_k
q^2}{q^2+m_{2k+1}^2}= \sum_k \frac{g^{(k)}_{V,mag}\zeta_k m_{2k+1}^2
}{q^2+m_{2k+1}^2}= \sum_k \frac{g_{v^{(k)}}
g^{(k)}_{V,mag}}{q^2+m_{2k+1}^2} \:.\label{chargeffm}
\end{equation}
Since the minimal term couples nucleons to $U(2)$ gauge field
and the magnetic term couples nucleons to $SU(2)$ gauge field,
the two form factors, $F_{1,min}$ and $F_{1,mag}$ contribute
differently to the proton and neutron charge form factors.

For this, note that both iso-scalars
and iso-vectors part of $v^{(k)}$ enter this cubic coupling,
unlike the case of pions
where only the iso-vector vectors enter the story.
The relative strength between the two are
determined universally by the 5D $U(N_F)$ charge of the baryon.
For $N_c=3$, $v$ in the baryon vertex is in the representation
\begin{equation}
v_\mu^{(k)}\simeq a^{(2k+1)}=\left(\begin{array}{cc}3/2 &0 \\ 0&3/2\end{array}\right)
 \omega^{(k)}_\mu+ \rho_\mu^{(k)} \:.
\end{equation}
The mixing between $v$ and ${\cal V}$ is computed
in the representation which is appropriate for mesons, where
\begin{equation}
v_\mu^{(k)}
\simeq \left(\begin{array}{cc}1/2 &0 \\ 0&1/2\end{array}\right)
 \omega^{(k)}_\mu+ \rho_\mu^{(k)}
\end{equation}
and
\begin{equation}
{\cal V}=\left(\left(\begin{array}{cc}1/6 &0 \\ 0&1/6\end{array}\right)+
\left(\begin{array}{cc}1/2 &0 \\ 0&-1/2\end{array}\right)\right){\cal V}_{em} \:.
\end{equation}
The representation of $v$ for nucleons  dictates the cubic coupling
of $v$ to the nucleon while the latter dictates the quadratic mixing
of $v$ and ${\cal V}$.

The electromagnetic interaction mediated by iso-scalars is thus
proportional to  $3/2\times (1/2\times 1/6)$ while its triplet
counterpart is proportional to $\pm 1/2\times (1/2\times 1/2)$
with the sign choice corresponding to choosing proton or neutron.
Since the two final products are equal in size,
iso-scalar vectors and iso-vector vectors contribute to the nucleon
form factor $F_{1,min}$ with the equal strength, adding up for the
proton and cancelling each other for the neutron. On the other hand,
only the iso-vector contribute to $F_{1,mag}$ with an opposite sign
for proton and neutron, respectively.
After taking into account the charge assignment for protons and
neutrons carefully, the electromagnetic charge form factors are
found as
\begin{eqnarray}
F_1^{proton}&=&F_{1,min}+\frac12 F_{1,mag}\:,\nonumber\\
F_1^{neutron}&=&-\frac12 F_{1,mag}\:.
\end{eqnarray}

\subsection{Pauli Form Factor $F_2$}

The phenomenon of complete vector dominance, that is, the absence of
direct coupling of photon to nucleons, is also seen in the Pauli
form factor $F_2(q^2)$ defined as
\be \langle B |J^\mu(q)|B
\rangle\sim {F^a_2(q^2)\over 2m_B} \bar B \gamma^{\mu\nu}p_\nu t_a B
\:. \ee
After inserting the mode expansion of the 5D gauge field
in terms of vector mesons into our bulk 5D magnetic coupling with
purely 4D polarizations, we obtain the interactions that are
relevant for magnetic dipole coupling, \be {g_2 \over 4m_B}\bar
B\gamma^{\mu\nu}{\cal F}_{\mu\nu} B+
 \sum_k {g_2^{(k)}\over 4m_B} \bar B \gamma^{\mu\nu}F_{\mu\nu}^{(2k+1)} B\:,
\ee
where $F_{\mu\nu}^{(2k+1)}=\partial_\mu a_\nu^{(2k+1)}-\partial_\nu a_\mu^{(2k+1)}$
is the "field strength" of the vector meson $a_\mu^{(2k+1)}$ and
${\cal F}_{\mu\nu}$ is the field strength of external source ${\cal V}_\mu$ for the current.
The coupling constants are easily read off from overlap integrals,
\bear
g_2&=& 0.18 N_c \times{4m_B\over M_{KK}}\times \int_{-w_{max}}^{w_{max}} dw f_L^*(w)f_R(w)\:,\nonumber\\
g_2^{(k)}&=&
0.18 N_c \times{4m_B\over M_{KK}} \times\int_{-w_{max}}^{w_{max}} dw f_L^*(w)f_R(w) \psi_{(2k+1)}(w)\:.
\eear
Note that contributions involving the axial vectors $a_\mu^{(2k)}$ are absent due to
their odd profile in the 5-th coordinate $\psi_{(2k)}(-w)=-\psi_{(2k)}(w)$ and the
property $f_L(-w)=f_R(w)$. In fact, the would-be terms like $\bar B \gamma^{\mu\nu}\gamma^5 F_{\mu\nu}^{(2k)} B
=\epsilon^{\mu\nu\alpha\beta}\bar B \gamma_{\mu\nu} F_{\alpha\beta}^{(2k)} B$ is CP-violating.

Using the completeness relation for $\psi_{(2k+1)}$ as before, it is
straightforward to check the sum rule \be \sum_k  g_2^{(k)}\zeta_k
=g_2\:, \ee which is saturated up to 99\% by the lowest four
vector mesons as can be seen in the Table 2. Because of this sum
rule, as we go to the shifted basis $v^{(k)}=a^{(2k+1)}+\zeta_k {\cal
V}$, the direct photon coupling ${g_2 \over 4m_B}\bar
B\gamma^{\mu\nu}{\cal F}_{\mu\nu} B$ is exactly cancelled by the
shift, and we are left with
\be
\sum_k {g_2^{(k)}\over 4m_B} \bar B
\gamma^{\mu\nu}(\partial_\mu v_\nu^{(k)}-\partial_\nu v_\mu^{(k)}
)B\:,
\ee
and the Pauli form factor is given as a sum over
intermediate vector meson contributions,
\be F^3_2(q^2)=\sum_k
{g_2^{(k)}\zeta_k m_{2k+1}^2\over q^2+m_{2k+1}^2}\:, \label{pauli_ff}\ee with
the property due to the sum rule $F^3_2(0)=g_2$. It seems by now clear
that the complete vector dominance is a generic phenomenon in the
holographic QCD, as was first noticed in~\cite{Hong:2004sa}.

For each nucleon, we have
\begin{eqnarray}
F_2^{proton}&=&\frac12 F_{2}^3\:,\nonumber\\
F_2^{neutron}&=&-\frac12 F_{2}^3\:,
\end{eqnarray}
since only the magnetic term contributes to $F_2$.

\subsection{Numerics and A Consistent Truncation}\label{section7_4}

It is tantalizing that there is a complete parallel between the
vector dominance in the pion and that in the nucleon.
Because the zero momentum limit of $F_1$ is the electromagnetic
charge, which should be quantized, $F_1(0)$ of pions and nucleons
must be the same, which would imply $g_{v^{(0)}\pi\pi} =
g^{(0)}_{V}$ if the sums were saturated by the lowest vector meson
$v^{(0)}$. This feature has been discussed much in old literatures
and goes by the name of ``universality."

To see whether the form factor is actually dominated by the first vector
meson or not, we computed numbers for the first few lowest vector
mesons and check the sum rules numerically. The result is shown
in table 2. We have two independent sum rules for
$g^{(k)}_{V, min}$ and for
sum $g^{(k)}_V$. Since the sum rules for $g^{(k)}_{V, min}$ and
$g^{(k)}_V$ are both tied to the net electromagnetic charge, we need to
satisfy them both well, before discussing any comparison with data.
As table 2 shows clearly, the sum rules that lead to the vector
dominance cannot be satisfied with the lowest vector meson alone,
indicating the truncation down to the first vector would be a bad
approximation for the form factor. Instead, if we sum up to the fourth
vector meson, both sum rules are obeyed with 0.2\% accuracy, giving
us a hope that $F_1$ may be well-approximated
in the low momentum region by summing over the first four terms.
A similar result can be seen for $F_2$,
since, as also shown in the table 2, the sum rule for $g_2$ is
saturated well within 1\% accuracy.

\begin{table}[h]
\begin{tabular}{|c|c|c|c|c|c|c|c|}
\hline $k$ & $m^2_{2k+1}$ & $\zeta_k$
& $g^{(k)}_{V,min}$ & $g_{V,mag}^{(k)}$ & $ g_{V,min}^{(k)}\zeta_k$
& $ g_{V,mag}^{(k)}\zeta_k$   & $g_2^{(k)}\zeta_k$  \\
 \hline\hline
0& 0.67 & 0.272 & 5.933 & -0.816 & 1.615 & -0.222 & 3.323  \\
 \hline
1 & 2.87 & -0.274 & 3.224 & -1.988 & -0.882& 0.544 & -1.918 \\
 \hline
2 & 6.59 & 0.272  & 1.261 & -1.932 & 0.343 & -0.526 & 0.828 \\
\hline
3 & 11.8 & -0.271 & 0.311 & -0.969 & -0.084 & 0.262 & -0.243 \\
\hline \hline {\rm sum} & - &- & -& -& 0.992 & 0.058  &1.989($g_2=2.028$)\\
\hline
\end{tabular}
\caption{\small  Numerical results for vector meson couplings for
the lowest four excitations in the case $\lambda N_c=50$. Sum
rules hold to a high precision. Our convention for the vector
meson fields differ by sign from that of Sakai and Sugimoto for
odd $k$. The vector meson mass squared is in the unit of
$M_{KK}^2$.} \label{table2}
\end{table}

Given this numerical data, the old "universality" seems to have
found a new reincarnation. The table 2 shows that the sum rules
for the nucleon (as well as for pions) are saturated within less
than 5\% by the four lowest vector mesons. We observe that
 \be
\zeta_k =(-1)^k /h
 \ee
where $h$ is a constant independent -- within less than 1\% -- of
the species $k$. Assuming that the sum rule (\ref{sumrule}) is
completely saturated by the four vector mesons, we arrive at the
conclusion that~\footnote{This is reminiscent of the nonet relation
in three flavor HLS$_1$
 \be
g_\rho/m_\rho^2=3g_\omega/m_\omega^2 =
-(3/\sqrt{2})g_\phi/m_\phi^2=1/g
 \ee
where $g$ is the hidden gauge coupling constant.}
 \be
\sum_{k=0}^3 (-1)^k g_{v^{(k)}\pi\pi}= h.\label{sumrule2}
 \ee
Since the same relation holds with a nucleon replacing the $\pi$ in
(\ref{sumrule2}), $h$ could be identified with the HLS$_\infty$
gauge coupling constant and that
 \be
 \sum_{k=0}^3 (-1)^k g_{v^{(k)}\pi\pi}\simeq\sum_{k=0}^3(-1)^k
 g_{v^{(k)}NN}.
 \ee
We could consider this as a ``generalized universality" relation,
although we have no rigorous argument for such a relation.

If the sum rules (\ref{sumrule}) and (\ref{crucialSR}) are saturated
by the first four vector mesons then an interesting question is how
the relation $\sum_{k=4}^\infty\zeta_k
g_{v^{(k)}\pi\pi}=\sum_{k=4}^\infty\zeta_k g_{v^{(k)}NN}=0$ is
satisfied and what it means vis-a-vis with the short-range structure
of the nucleon. We leave these issues for later publication.

\subsection{The ``Old" Vector Dominance in Light of the ``New" Vector Dominance}
Now that we have the form factors of both pions and nucleons
completely vector-dominated with the infinite tower, it is
interesting to review the old vector dominance involving the lowest
vector mesons only, $\rho$, $\omega$ and $\phi$ -- and in some works
including the next-lying vector mesons~\cite{lomon} -- in light of
the new picture. This could bring light to the success and failure
of the old vector dominance. We shall do this using the
Harada-Yamawaki (HY) approach~\cite{HY:PR}.

As has been suggested~\cite{haradaetal-ads-hls}, HY's hidden local
symmetry model can be considered as resulting from integrating out
all excitations other than the pions and the lowest vector mesons
and matching the truncated action to the SS action at a matching
scale $\Lambda_M$. It is however more natural to consider it as an
emergent symmetry as mentioned in Section \ref{2.2}. It is in this
way via what is called ``moose construction"~\cite{georgi} that the
tower of vector mesons emerge in a dimensionally deconstructed
QCD~\cite{son-stephanov} with a five-dimensional YM action analogous
to that reduced from string theory that we have been discussing.

What we would like to do here is to describe how the vector
dominance and the putative violation thereof arise in this HLS
approach (that will be referred to as HLS/VM below). In order to do
so, we recall how massive vector meson degrees of freedom arise when
one approaches hadron chiral dynamics from bottom up. At very low
energy, $E\ll \Lambda_\chi$, i.e., the chiral scale, the chiral
dynamics is given by the current algebra term with the
lowest-derivative Lagrangian
 \be
{\cal L}=\frac{f_\pi^2}{4}\Tr (\del_\mu U\del^\mu
U^\dagger)\label{current algebra}
 \ee
with the chiral field
 \be
U(x)=e^{2i\pi/f_\pi}.
 \ee
By writing the $U$ field as a product field
 \be
U=\xi_L^\dagger \xi_R
 \ee
which can be done by introducing a redundant field $\sigma$ as
 \be
\xi_{L/R}=e^{\mp i\pi/f_\pi}e^{i\sigma/f_\sigma}
 \ee
with $f_\sigma$ defined as the $\sigma$ decay constant, one unearths
a trivial local invariance
 \be
\xi_{L/R}\rightarrow h(x)\xi_{L/R}\label{xi}
 \ee
with $h(x)\in U(N_F)$. This local symmetry can be exploited by
introducing a vector field $v_\mu$ to bring the energy scale from
low, here that of the pion mass -- which is zero in the chiral limit
with the Lagrangian (\ref{current algebra}) -- to high, say, the
scale set by the mass of a meson $v$. This is essentially how the
vector mesons $(\rho,\omega)$ were incorporated into the HLS theory
of \cite{bandoetal}. For convenience, we shall call it HLS$_1$.

As is well-known~\cite{weinberg-gaugetheory}, the gauge theory so
constructed does not lead to a unique higher-energy theory. In order
to direct the hidden gauge theory of \cite{bandoetal} toward a
correct one, Harada and Yamawaki match \`a la Wilson the effective
theory to QCD at a matching scale $\Lambda_M\sim 4\pi f_\pi$.
Specifically the vector correlator $\Pi_V$ and the axial-vector
correlator $\Pi_A$ calculated with the HLS Lagrangian are matched to
those calculated in QCD, e.g., operator-product expansion (OPE).
This allows the $bare$ parameters of the HLS Lagrangian, $g$,
$f_\pi$ and $f_\sigma$, to be expressed in terms of the QCD
variables, $\alpha_s$, $\la\bar{q}q\ra$, $\la G^2\ra$ etc. Given the
bare Lagrangian, the next step is to do renormalization group
analysis to see how the theory flows as the scale is changed from
the matching scale. Harada and Yamawaki find a variety of fixed
points as well as a fixed line, to which the HLS$_1$ can
flow~\cite{HY:PR}. In order to pick out the fixed point that maps to
QCD, one has to impose the condition that when the chiral order
parameter $\la\bar{q}q\ra$ is set equal to zero, the correlators are
equal, i.e., $\Pi_V=\Pi_A$. This condition picks out the fixed point
that corresponds to the fixed point to which the system flows when
the condensate $\la\bar{q}q\ra$ goes to zero. This fixed point
called ``vector manifestation (VM) fixed point" (and the HLS theory
with the VM fixed point called HLS/VM)\cite{HY:PR} is characterized
by
 \be
g^*=0, \ \ a^*=1\label{VM}
 \ee
where $a$ is the ratio of the decay constants $a\equiv
(f_\sigma/f_\pi)^2$. This fixed point is reached when a hadronic
system in Nambu-Goldstone phase makes the transition to the symmetry
restored phase at high temperature $T_c$ (as in the Early Universe)
or at high density $n_c$ (as in compact stars).

What this implies in the EM form factors of the pion and the nucleon
in HLS$_1$ is as follows. In HLS theory with the lowest vector
mesons $(\rho,\omega)$, the iso-vector photon coupling is given by
 \be
\delta {\cal L}=e{\cal A}_{EM}^\mu\left(-2af_\pi^2  {\rm
Tr}[g\rho_\mu Q] +2i(1-a/2) {\rm Tr}[J_\mu
Q]\right)\:,\label{coupling})
 \ee
where $Q$ is the quark charge matrix, $\rho_\mu$ is the lowest-lying
iso-vector vector meson and $J_\mu$ is the iso-vector vector current
made up of the chiral field $\xi$ (\ref{xi}). The first term of
(\ref{coupling}) represents the photon coupling through a $\rho$ and
the second term the direct coupling. The ``old" vector dominance is
obtained when $a=2$ for which the well-known KSRF relation for the
$\rho$ meson holds, i.e., $m_\rho^2=af_\pi^2 g^2=2f_\pi^2 g^2$. Now
it has been established empirically that the way the vector
dominance manifests itself is different between the pion and the
nucleon. Let us look at them separately.
 \begin{itemize}
\item Pion form factor: On-shell in matter-free space, the pion form
factor is very well described by the vector dominance, hence $a=2$,
with no direct coupling. However in HLS/VM, in the framework of
HLS$_1$, $a=2$ is totally accidental, not even lying on a stable
trajectory of the RGE~\cite{HY:PR,HY:VD}. A small perturbation would
take $a$ away from the vector dominance point $a=2$. Thus for
instance, temperature~\cite{HS:T} or density would push $a$ toward
1, inducing what is referred to as ``vector dominance violation." It
is an interesting possibility that there is a connection, albeit
indirect, between the departure from $a=2$ toward $a=1$ in HLS$_1$
and the role in medium of higher-lying vector mesons in
HLS$_\infty$. This is an important issue in CERN experiments on
dilepton production in relativistic heavy ion
collisions~\cite{HS:dilepton}.
\item Nucleon form factor: If one considers nucleon as a Skyrmion in
HLS$_1$, then the second term in (\ref{coupling}) corresponds to a
direct photon coupling to the Skyrmion. As we will elaborate in
Section 9, experimental data clearly show that there is an important
direct coupling with $a\sim 1$. This observation has been taken as
an indication that vector dominance does not apply to nucleons, the
reason put forward for this violation being that nucleons are
extended objects. We will see in Section 9 that this picture is
drastically modified when the infinite tower of vector mesons enter
in the structure of nucleons.
 \end{itemize}
\section{The Anomalous Magnetic Dipole Moment\label{MM}}

While we can simply read out the magnetic moment from the form
factors of previous section, here we would like to show a more direct
computation, which does not depend on the structure of the
vector dominance of the previous section. Since the magnetic
moment comes from form factors at zero momentum
limit, it is best to work in $\{{\cal V};a^{(2k+1)}\}$ basis and ignore
the vector mesons entirely. This is because $a^{(2k+1)}$'s mixes with
${\cal V}$ at the level of kinetic term and thus with two momentum factors
while $v^{(k)}$ mixes with ${\cal V}$ at the mass term level.
Thus, we may ask how the nonnormalizable mode
${\cal V}$ in
\begin{equation}
A_\mu(x;w)={\cal V}_\mu(x)+i\alpha_\mu(x)\psi_0(w)+\cdots
\end{equation}
couples to the nucleons. Let us insert the
non-normalizable zero mode into the effective action for the five
dimensional baryon, whereby we find the terms relevant
\begin{eqnarray}
\int d^4 x \int dw\left[ -\bar{\cal B}\gamma^{\mu}  {\cal V}_\mu' {\cal B}
 +g_5(w){\rho_{baryon}^2\over
e^2(w)}\bar{\cal B}\gamma^{\mu\nu}{\cal F}_{\mu\nu}{\cal B} \right]
\end{eqnarray}
with $g_5(0)=2\pi^2/3$. Here we denoted the gauge field from the
minimal coupling by ${\cal V}'$ because its generator is different
from the one in the magnetic term.

Again recall that the 5D magnetic coupling that we obtained from comparing
with long-range instanton tail must contain only the $SU(2)$ isospin
without an overall $U(1)$, since the instanton tail involves only
non-Abelian $SU(2)$. On the contrary, the minimal coupling term
contains $U(1)_Y$ as well as $SU(2)$ according to the charge of
nucleons made out of $N_c$-quarks. As for the case $N_c=3$ and
$N_F=2$, the quark doublet $(u,d)$ has EM charge $(2/3,-1/3)$, which
can be decomposed to $(1/6,1/6)$ corresponding to $U(1)_Y$ and
$(1/2,-1/2)$ for the diagonal part of $SU(2)$. As nucleons are made
of 3 quarks in totally anti-symmetric fashion, the resulting
$U(1)_Y$ charge becomes $(1/2,1/2)$ whereas the $SU(2)$ charge
remains fundamental representation $(1/2,-1/2)$. This tells us that
we have to use EM charge $(1,0)$ for $(p,n)$ in the minimal coupling
as expected, while we should instead have $(1/2,-1/2)$ in the term
from 5D magnetic coupling.

The above descends down to similar 4D expression as
\begin{eqnarray}
\int d^4 x \left[ -\bar{B}\gamma^{\mu}  {\cal V}_\mu' {B}
 +g_5(0){\rho_{baryon}^2\over
e^2(0)}\bar{B}\gamma^{\mu\nu}{\cal F}_{\mu\nu}{B} \right] \:,
\end{eqnarray}
assuming that the  eigenmode $f_{L,R}$ of the nucleon is
sufficiently concentrated at origin $w=0$, so that the
$w$-dependence of the coupling does not enter the physics. The
previous estimate gives us
\begin{equation}
g_5(0){\rho_{baryon}^2\over
e^2(0)}\simeq 0.18N_c \times{1\over M_{KK}}\:.
\end{equation}
This approximation becomes precise in the large $N_c$-limit. For
later numerical calculations extrapolating to $N_c=3$, the precise
overlap integral replaces the above coefficient with
\begin{equation}
0.18N_c \times{1\over M_{KK}}\times \int_{-w_{max}}^{w_{max}}
dw\, {g_5(w)U(w)\over g_5(0)U_{KK}}f_L^*(w)f_R(w)\:,
\label{magmom}
\end{equation}
where we used $f_L(w)=f_R(-w)$ for the lowest nucleon eigenmode.

Taking a non-relativistic limit, we will look for terms of type
\begin{equation}
\int d^{4}x\;\frac{\mu}{e_{EM}}\,{\bf S}\cdot {\bf B}
\end{equation}
with the magnetic field strength ${\cal B}$ and the spin ${\bf S}$.
Here we have a factor of $e_{EM}$ to correct the fact that our
choice of gauge field is not canonically normalized. As in section
(\ref{5DB}), we introduce the two-component notation of $B$ as
\begin{equation}
B=\left(\begin{array}{c} u\\ v\end{array}\right) e^{-iEt+i p\cdot  x}\:,
\end{equation}
where the on-shell condition relates
\begin{equation}
v=\frac{E-\sigma\cdot p}{-im_B}\,u\:.
\end{equation}
Isolating the magnetic dipole coupling, we find
\begin{equation}
\frac{1}{m_B}\int d^4x\; \left[u^\dagger {\bf B}'\cdot \sigma u
\right]+ \left(\frac{4g_5(0)\rho^2_{baryon}}{e(0)^2}\right)\int
d^4x\; \left[u^\dagger {\bf B}\cdot \sigma u \right]\:.
\end{equation}
where, again, the prime on the magnetic field  reminds us
that the charge generator for the minimal coupling term is
different from the one for the magnetic term.
Given the normalization, $\tr u^\dagger  u=1/2$
(see section.~\ref{5DB}), one can identify $\tr u^\dagger \sigma u$
as the spin operator ${\bf S}$ of the nucleon. This leads to
\begin{equation}
\frac{\mu_{proton}}{e_{EM}}=
\frac{1}{2m_B}+\left[\frac{g_5(0)\rho^2_{baryon}}{e(0)^2}\right],\qquad
\frac{\mu_{neutron}}{e_{EM}}=-\left[\frac{g_5(0)\rho^2_{baryon}}{e(0)^2}\right]\: .
\end{equation}

However, we do not have a reliable estimate of the nucleon mass
$m_B$. One way to bypass this difficulty is to look at the anomalous
part of the magnetic dipole moment. In fact, the anomalous part is
the dominant part in the large $N_c$ limit, and thus is likely to be
more reliable. We have
\begin{equation}
\frac{\mu_{proton}^{an}}{e_{EM}}=\frac{0.18N_c}{M_{KK}}\:,\qquad
\frac{\mu_{neutron}^{an}}{e_{EM}}=-\frac{0.18N_c}{M_{KK}}\:.
\end{equation}
For comparison with experiments, let us first consider the
difference of the anomalous magnetic moment $\Delta
\mu^{an}=\mu^{an}_{proton}-\mu^{an}_{neutron}$,
\begin{equation}
\frac{\Delta \mu^{an}}{e_{EM}}
\simeq  {0.36N_c\over M_{KK}}\:.
\end{equation}
Experimentally, $(\Delta\mu^{an})_{exp}=(2.79-(-1.91)-1)\times
\mu_N=3.7 \mu_N$ where $\mu_N={e_{EM}/ 2 m_N}$ is the nuclear
magneton. Once we take $M_{KK}=0.94{\rm GeV}$ as determined by the
meson sector fit, it happens to be approximately the physical
nucleon mass, denoted as $m_N$. Thus our prediction is
$\Delta\mu^{an}\simeq 0.72N_c \times \mu_N=2.16 \mu_N$ for $N_c=3$.

However, if we replace $N_c$ by $(N_c+2)$ again guided by CQM, then it
becomes
\begin{equation}
\Delta\mu^{an}\simeq 3.6\mu_N \:,
\end{equation}
which agrees with experiment value, $3.7\mu_N$, very well.  With
the same shift, the individual anomalous magnetic moment are
\begin{equation}
\mu_{proton}^{an}\simeq 1.8\mu_N\:,\qquad
\mu_{neutron}^{an}\simeq -1.8\mu_N:,
\end{equation}
which again compare quite favorably to the experimental values,
$1.79\mu_N$ and $-1.91\mu_N$, respectively. Such a shift $N_c
\rightarrow N_c+2$ was discussed in Section \ref{shift} for the
leading chiral coupling between the pion and the nucleon. As
mentioned there, the spin-isospin structure is the same for the
axial coupling and the iso-vector magnetic moment, so the
collective quantization leads to the same shifting for both. 

A thorny issue here, and also for much of next section where
we consider electromagnetic form factors, is the matter of the
nucleon mass $m_B$. For instance, the non-anomalous part of the
proton magnetic moment would be computed to be $e_{EM}/2m_B$
and the question of whether the model predicts $m_B\simeq m_N$
becomes an important issue.

In this article we did not attempt to compute $m_B$ within our
model. In fact, it is unclear if there should exist an unambiguous
prediction for the ground state mass in this approach, since the
quantity is additively renormalized, since an infinitely many
oscillators around the classical soliton contribute zero-point
energy. Hata et.al. \cite{Hata:2007mb} computed the mass spectra
of various excited baryons but, for this reason, chose to the
treat the ground state mass ($m_B$ in our notation) as a free
parameter instead. For a bona fide comparison of quantities that
depends on the nucleon mass sensitively, this issue should be
resolved first.

\section{Electromagnetic Form Factors}
\subsection{Two-Component Description}
The full electromagnetic form factors are encoded in three
functions, $F_{1,2,3}$. Before we compute the form factors to
compare with the experimental data, we review briefly the past
theoretical status on the subject.

For qualitative illustration of what the problem is, we take the
iso-vector Dirac form factor $F_1$ of the nucleon. This form factor
will receive contributions from the vector mesons in the tower in
the $\rho$ channel, $\rho$, $\rho^\prime$, etc. Other channels can
be discussed in a similar way.

In the literature, analysis have been made by including
one~\cite{bijker-iachello} or two~\cite{lomon} lowest vector
mesons in the $\rho$ channel, i.e., $\rho (770)$ and $\rho^\prime
(1450)$. Let us just take the lowest only for the discussion,
relegating the role of $\rho^\prime$ to a short comment later.
\begin{figure}[ht]
\begin{center}
{\includegraphics{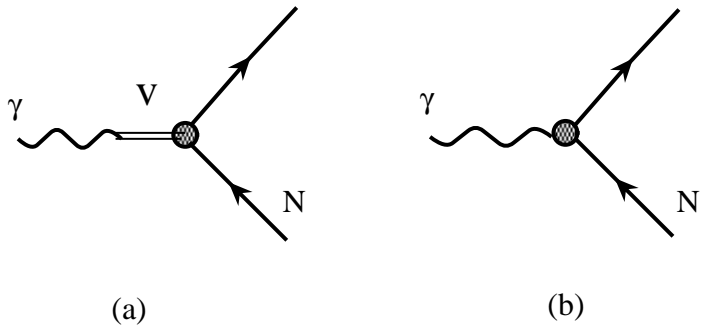}}
\end{center}
\vskip -7cm\caption{\label{photoncoupling} (a) Photon coupling to
the nucleon via vector meson $V$ and (b) direct photon coupling to
the nucleon. The blob represents the intrinsic form factor
accounting for short-distance effects (referred to as ``intrinsic
core" in some circles) unaccounted for in the effective theory,
e.g., asymptotically free QCD property.}\label{VD-Nuc}
\end{figure}

It has been known since a long time that the nucleon form factors at
low momentum transfers cannot be fitted by a monopole form factor of
the type $\sim 1/(1+cq^2)$ with $c$ a constant where $q$ is
Euclidean four momentum transfer. In fact, one obtains a much better
fit by a dipole form factor of the form $\sim 1/(1+dq^2)^2$ with
$d\approx 1/(0.71\ {\rm GeV})^2$. This meant that the
single-vector-meson mediated mechanism along the line of reasoning
used for the pion form factors could not explain the process. This
led to a two-component description Fig.\ref{VD-Nuc} which can be put
in the form~\cite{iachello-jackson-lande}
 \be
F_1 (q^2)=\frac
12\left[A(q^2)+B(q^2)\frac{m_\rho^2}{q^2+m_\rho^2}\right]\label{IJL}
 \ee
with the normalization
 \be
A(0)+B(0)=1.
 \ee
In (\ref{IJL}), the momentum dependence in $A$ and $B$ represents a
form-factor effect corresponding to an intrinsic structure of the
nucleon which is expected from both the confinement and the
asymptotic behavior of perturbative QCD~\cite{lepage-brotsky}. The
first term corresponds to a direct coupling to the intrinsic
component of the nucleon (``nucleon core" in short),
Fig.\ref{VD-Nuc}b and the second term via one or more vector mesons
in the $\rho$ channel, Fig.\ref{VD-Nuc}a. Let us for simplicity
consider only one vector meson exchange. The same reasoning applies
to the case where more than one vector mesons are considered. Making
the reasonable assumption that the photon and the $\rho$ meson
couple to the nucleon core with a same form factor, one can rewrite
(\ref{IJL})  as
 \be
F_1 (q^2)=\frac 12 h(q^2) \left[(1-\beta_\rho) + \beta_\rho
\frac{m_\rho^2}{q^2+m_\rho^2}\right]\label{IJL-BI}
 \ee
with the core form factor normalized as $h(0)=1$. Perturbative QCD
indicates that asymptotically $h(q^2) \approx (1+\gamma q^2)^{-2}$.
The coefficient $\gamma$ is not given by the model but can be fixed
by experimental data. One can make a very good fit to the data with
(\ref{IJL-BI}) with the coefficients $\gamma\approx 0.52\ {\rm
GeV}^{-2}$ and $\beta_\rho\approx 0.51$~\cite{bijker-iachello}.

Let us consider what this result means with regard to our prediction
of Section (\ref{VD-N}). The VD prediction (\ref{chargeff}), as
mentioned above, lacks the intrinsic short-distance form factor but
this can be implemented, albeit phenomenologically as in the
two-component model, since it involves physics intervening at a
scale above the KK scale. What is significant is the role of the
first term of (\ref{IJL-BI}). In the two-component model, this part,
characterized by a size of $\sim 0.4$ fm, is to represent the
short-distance physics of the microscopic degrees of freedom of QCD
that are extraneous to long-wavelength excitations -- $\pi$, $\rho$
etc. -- in the baryon. Adding the $\rho^\prime$ meson and higher in
the second term of (\ref{IJL-BI}) is expected to further reduce the
size of the core. One interpretation of the core component was given
in terms of a ``chiral bag" in which quarks and gluons are confined
with the broken chiral symmetry of QCD suitably implemented outside
of the bag~\cite{BRW-VD}. The baryon charge was assumed to be
divided roughly half and half between the quark-gluon sector and the
hadron sector. This hybrid model met with a fair success in
reproducing the data available up to late 1980's~\cite{BRW-VD}.
Interestingly, it has been claimed that there is an (albeit
indirect) evidence for a core of $\sim 0.2$ fm from the Nachtmann
moment of the unpolarized proton structure function measured at
JLab~\cite{petronzio}.

Within the framework of the two-component picture, an alternative
description using the Skyrmion as an extended object to which the
photon couples both directly and via the exchange of the lowest
member of the vector-meson tower has been
constructed~\cite{holzwarth}. With one parameter that represents the
amount of direct coupling, the model is found to agree quite well
with the dipole form factors up to $q^2\sim 1$ GeV$^2$ and can
explain satisfactorily the deviation from the dipole form for
$q^2\gsim 1$ GeV$^2$. What this implies is that the nucleon form
factors at low momentum transfers, say, $q^2\lsim 1$ GeV$^2$, can be
well understood given the three basic ingredients: (a) an extended
object, (b)partial coupling to vector mesons and (c) relativistic
recoil corrections.

What we have found in the holographic dual model in Section
\ref{VD-N} is that by a suitable field re-definition and using a sum
rule involving the spread in the fifth dimension, one can transform
away the ``contact" coupling Fig.\ref{VD-Nuc}b -- here to the
soliton -- at the expense of saturating Fig.\ref{VD-Nuc}a with the
infinite tower of the vector mesons. The novel structure of this
model is that the ``intrinsic core" is largely replaced by the
higher-lying vector mesons in the infinite tower encapsulated in the
instanton baryon -- modulo the asymptotically free property relevant
at very high momentum transfer not captured in the model, say,
physics of $\lsim$ 0.2 fm. We will see indeed that this small core
size is needed for phenomenology.

\subsection{Instanton Baryon Prediction for the Form Factors}
The nucleon form factors are defined from the matrix elements of the
external current operator $J^{\mu}$ as
\begin{equation}
\left<p^{\prime}\right|J^{\mu}(x)\left|p\right>=e^{iqx}\,\bar
u(p^{\prime})\,{\cal O}^{\mu}(p,p^{\prime})\,u(p) \:,
\end{equation}
where $q=p^{\prime}-p$. By the Lorentz invariance and the current
conservation we may expand the operator ${\cal O}^{\mu}$ as
\begin{equation}
{\cal
O}^{\mu}(p,p^{\prime})=\gamma^{\mu}\left[\frac12F_1(q^2)+F^a_1(q^2)\tau^a\right]
+\frac{\gamma^{\mu\nu}}{2m_B}q_{\nu}
\left[F_2(q^2)+F^a_2(q^2)\tau^a\right]\:,
\end{equation}
where $F_1$ and $F_2$ are the Dirac and Pauli form factors
for iso-scalar current respectively, and  $F_1^a$, $F_2^a$ are for iso-vector currents.
Our convention is $\tau^a=\sigma^a/2$.

In the AdS/CFT correspondence the matrix element is given by the
overlap integral of the normalizable modes, corresponding to the
nucleon states, and a non-normalizable mode of gauge fields
$A_{\mu}(x,z)$, which becomes an external source for the current
at the UV boundary. By matching the correct operators from the 5D
effective action in Eq.~(\ref{5dfermion1}), one can read off the
corresponding form factors.

We first Fourier-transform the gauge fields of the external source
of currents as
\begin{equation}
A_{\mu}(x,z)=\int_q\,A_{\mu}(q)A(q,z)\,e^{iqx}\:.
\end{equation}
{}From the equation of motion for the gauge field we get
\begin{equation}
\left(1+z^2\right)^{4/3}\partial_z^2\,A(q,z)+2z\left(1+z^2\right)^{1/3}
\partial_z\,A(q,z)-q^2\,A(q,z)=0
\end{equation}
with boundary conditions for all $q$
\begin{equation}
\lim_{z\to\pm\infty}A(z,q)=1,\quad
\lim_{z\to\pm\infty}\partial_zA(q,z)=0\:.
\end{equation}
After solving this and inserting it into our 5D action, we can
read off suitable form factors at momentum $q^2$. We note that the
Dirac form factor is a sum of a term, $F_{1\rm min}$, coming from
the minimal coupling and a term, $F_{1\rm mag}$, coming from the
magnetic coupling, which are
\begin{eqnarray}
F_{1{\rm min}}(q^2)&=&\int_{-w_{max}}^{w_{max}} dw\,\left|f_L(w)\right|^2\,A(q,z(w))\:,\\
F_{1\rm mag}(q^2)&=&2\times 0.18N_c\int_{-w_{max}}^{w_{max}}
dw \left(g_5(w)U(w)\over g_5(0)U_{KK}M_{KK}\right)\left|f_L(w)\right|^2\partial_w A(q,z(w))\:.\nonumber
\label{ff_1}
\end{eqnarray}
where $f_{L,R}(z)$ are the left(right)-handed normalizable modes,
corresponding to the nucleon state. The Pauli form factor is given
as
\begin{eqnarray}
F^3_{2}(q^2)&=&0.18N_c\times\frac{4m_B}{M_{KK}}\,
\int_{-w_{max}}^{w_{max}} dw\,
\frac{g_5(w)U(w)}{g_5(0)U_{KK}}f_L^*(w)f_R(w)\,A(q,z(w))\:.
\label{ff_2}\end{eqnarray} One salient prediction of instanton
baryons on the form factor is that the $U(1)$ part of the Pauli form
factor $F_2(q^2)=0$, because the instanton does not have a $U(1)$
tail, while $F_{1\rm min}^3(q^2)=F_{1\rm min}(q^2)$. We also note
that our expressions for the form factors are from the AdS/CFT
correspondence, for which we have to use full 5D effective action
rather than using the leading two terms in the derivative expansion.
Therefore, our results cannot be trusted for $q^2\gsim M^2_{\rm
KK}$\:.

\begin{figure}[ht]
\centerline{\includegraphics[scale=1.0]{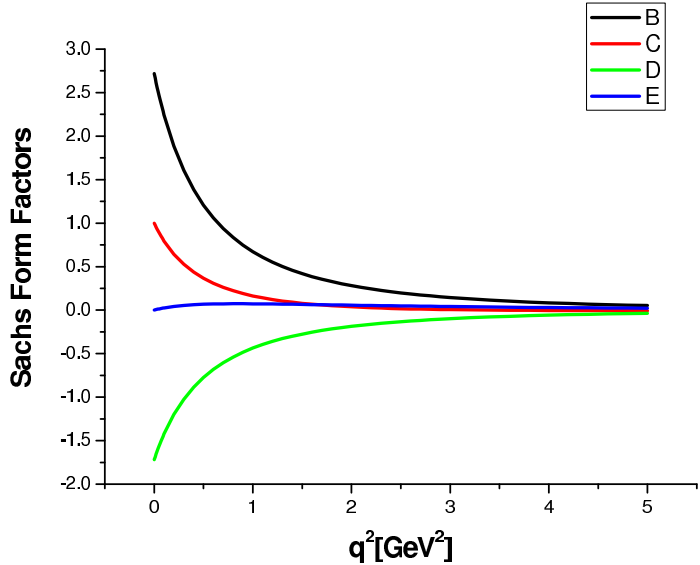}}%
\caption{The Sachs form factors vs. $q^2$ in ${\rm GeV}^2$:
B=$G_M^p$, C=$G_E^p$, D=$G_M^n$, and E=$G_E^n$, where we take
$m_B=M_{KK}$ and have shifted $N_C\to N_C+2$.}
 \label{fig3}
\end{figure}

The experimentally measured nucleon form factors (Sachs form
factors)  are defined for the space like momentum transfer, $q^2>0$,
as
\begin{eqnarray}
G_M^p(q^2)&=&F_{1\rm min}(q^2)+{1\over 2}F_{1\rm mag}(q^2)+{1\over 2}F_2^3(q^2):,\\
G_E^p(q^2)&=&F_{1\rm min}(q^2)+{1\over 2}F_{1\rm
mag}(q^2)-\frac{q^2}{4m_B^2}{1\over 2}F_2^3(q^2)\:,\\
G_M^n(q^2)&=&-{1\over 2}F_{1\rm mag}(q^2)-{1\over 2}F_2^3(q^2)\:,\\
G_E^n(q^2)&=&-{1\over 2}F_{1\rm mag}(q^2)+\frac{q^2}{4m_B^2}{1\over 2}F_2^3(q^2)\:.
\end{eqnarray}
For the numerical analysis we need to know the coordinate dependence
of the magnetic coupling $g_5(w)/e^2(w)$, which is for simplicity
approximated as $g_5(w)/e^2(w)\simeq g_5(0)/e^2(0)$.

Our results are plotted in Fig.~\ref{fig3}. To meaningfully compare
our results with experiments, there are several corrections to be
taken into account that are left out in our theory. One of the most
important of them  that influences the iso-vector from factors at
low momentum transfer is that the lowest iso-vector vector meson
$\rho$ has a large width, $\sim$ 150 MeV, which in our treatment
corresponds to higher order in $1/N_c$ expansion and hence is
absent. As mentioned above, the short-distance physics involving a
scale higher than the KK mass $M_{KK}$ given in QCD as an asymptotic
scaling~\cite{brotskyetal} is also missing. It is therefore with
these caveats in mind that our results for the Sachs form factors
given in Fig.~\ref{fig3} should be viewed. To have an idea as to how
they fare with Nature, let us look at the first nontrivial moment of
the proton form factors, namely, $dG^p(q^2)/dq^2|_{q^2=0}$
corresponding to charge (magnetic) square radius. For very low
momentum transfers, $q^2\ll 1$ GeV$^2$, the form factors can be
written as
 \be
G^p (q^2)\approx 1-\frac 16\la r^2\ra q^2 +\cdots\:,
 \ee
Our results of Fig.~\ref{fig3} give
 \be
\sqrt{\la r^2\ra_E^p}\simeq 0.80\ {\rm fm}, \ \  \sqrt{{\la
r^2\ra}_M^p}\simeq 0.74\ {\rm fm}.
 \ee
The empirical values~\cite{belushkinetal06} determined from
experiments via dispersion relation analysis are
 \be \sqrt{\la
r^2\ra_E^p}=0.886\ {\rm fm}, \ \  \sqrt{{\la r^2\ra}_M^p}=0.855\
{\rm fm}.
 \ee
By comparing the predictions with the empirical results, we can note
that the predicted sizes -- both electric and magnetic -- are
smaller than the experimental sizes by $\sim 0.15-0.17$ fm, roughly
the size of the ``intrinsic core" seen in inelastic electron
scattering experiments~\cite{petronzio}. Since the radii are
smaller, the form factors are expected to fall more slowly than
observed at low momentum transfers. However what is significant is
that the deviations are of the same magnitude, i.e., $\sim
0.15-0.17$ fm for both charge and magnetic radii. That they come out
to be the same can be understood by that the ``core" reflects
short-distance physics more or less ``blind" to flavor and spin.
This suggests that the ``core" effect should cancel out in the ratio
$R_p\equiv \frac{\mu_p G_E^p}{G_M^p}$. It indeed does. The predicted
value at $q^2=0.1$ GeV$^2$ is
 \be
R_p (q^2=0.1 {\rm GeV}^2)\approx 0.966
 \ee
to be compared with the empirical value
 \be
R_p (q^2=0.1 {\rm GeV}^2)\approx 0.97.
 \ee

Another way of calculating the form factors is to expand the
non-normalizable mode in terms of the normalizable modes, $\psi_{(2k+1)}(z)$,
of vector mesons in the overlap integrations~(\ref{ff_1})
and~(\ref{ff_2}),
\begin{equation}
A(q,z)=\sum_k\frac{g_{v^{(k)}}\psi_{(2k+1)}(z)}{q^2+m_{2k+1}^2}\:,
\label{expansion}
\end{equation}
where $m_{2k+1}$ and $g_{v^{(k)}}$ are the mass and  the decay
constant of the $k$-th vector mesons~\cite{Hong:2004sa}. (Note that
the axial vector mesons should enter to form a complete set when we
expand the non-normalizable mode. However, since the overlap
integration for the Dirac and Pauli form factors is parity even
under the parity flip of the 5th coordinate, the axial vectors do
not contribute.) Then we will get the previously defined form
factors Eq's ~(\ref{chargeff}) and (\ref{pauli_ff}), where the
vector meson decay constant is given by
\begin{equation}
g_{v^{(k)}}=\zeta_k\: m_{2k+1}^2\:.
\end{equation}
This shows that as noted in~\cite{Hong:2004sa}, the vector dominance
in the form factors for both the pion and the nucleon is a direct
consequence of AdS/CFT.

To illustrate that the vector-dominance description captures the
same physics as the instanton picture, we calculate the iso-vector
charge radius (ICR) of the proton by saturating the charge form
factor by the four lowest vector mesons in the $\rho$ channel.
Numerically, $\zeta_k$ are a constant. We take $\zeta=0.27$ and find
from Table \ref{table2}
 \be
\sqrt{\la
r^2\ra^p_C}\simeq\left(6\zeta\sum_{k=0}^3\frac{g_V^{(k)}}{m_{v^k}^2}{\rm
sign}\zeta_k\right)^{1/2}\simeq 0.63\ {\rm fm}.\label{ICR}
 \ee
The ``empirical value" represented by the dipole parametrization
$1/(1+Q^2/m_V^2)^2$ with $m_V=0.84$ MeV is $\sqrt{\la
r^2\ra^p_C}=0.81$ fm, so we find the predicted charge radius is
smaller than the empirical one by $\sim 0.18$ fm, about the same
as what we found with the Sachs form factors.

\section{Summary and Comments}

In this article, we pursued an holographic realization of baryons in
the SS model of QCD. In this model, the entire meson sector of
quenched QCD is collectively realized as a five-dimensional $U(N_F)$
gauge field and the KK tower produced upon a dimensional reduction
gives the towers of vector mesons and axial vector mesons, while an
open Wilson line corresponds to the chiral field of pions, $U$. The
string theory, in which the model is embedded, tells us
unambiguously that the baryon arises by quantizing an instanton
soliton of $SU(N_F)$ gauge field in five dimensions. We studied its
static property in large $\lambda$ and large $N_c$ limit, as
demanded by the classical approximation on the bulk side to the
AdS/CFT correspondence, and found the soliton size scales as $\sim
1/(M_{KK}\sqrt{\lambda})$. The small size motivates us to set up an
effective action approach treating the soliton as a point-like
object, and we explored its consequence in detail. The picture that
arises is consistent with heavy-baryon chiral effective field theory
where baryons are taken as local fields, with higher order
corrections in derivative and/or $1/N_c$ expansion, accounting for
the finite size of the baryons.

One might wonder how this small instanton soliton is related to
the usual Skyrmion in the four dimensional chiral lagrangian
approach~\footnote{Here and in what follows, by ``usual Skyrmion,"
we mean the Skyrmion arising from the Skyrme Lagrangian consisting
of the pion hedgehog. This should be distinguished from the
Skyrmion involving the infinite tower of vector mesons that
emerges from the 5D instanton.} in which the size of the baryon is
mostly given by the soliton size. Both objects are classified by
the topological charge $\pi_3(SU(N_F=2))$, and the topological
relation can be made precise by the following mapping \cite{Atiyah:1989dq}: Let $A$ be
an instanton in $R^3\times I$ with the unit Pontryagin number.
Then the open Wilson line
\be U=Pe^{i\int_I A}\:, \ee
 as a
function $R^3\rightarrow SU(2)$, carries a unit winding number in
$\pi_3(SU(2))$. The latter is of course the definition of the
Skyrmion winding number. Topologically this shows why the
instanton soliton is the underlying five-dimensional object which
produces the Skyrmion upon dimensional reduction to four
dimensions.

However, the question of size must be addressed. The Skyrmion
solution that would have come out of the chiral lagrangian is of
size $\sim 1/M_{KK}$. Yet, the instanton soliton we found has the
size $\sim 1/(M_{KK}\sqrt{\lambda})$, which is much smaller when
compared to the expected Skyrmion size. Upon the above map from the
instanton soliton to Skyrmion, we can see also that the size of the
latter essentially is the size of the former. So what went wrong?
The answer is that the usual Skyrmion is a bad approximation to the
baryon once we begin to include massive vector and axial-vector
mesons. Likewise, the truncation down to the usual chiral Lagrangian
involving only the pion field is also a bad approximation once we
begin to consider the baryonic sector of QCD as previously
suspected~\cite{brihayeetal,rho}.

{}From the four dimensional viewpoint, one can understand this
disparity of sizes by incorporating more and more of massive vector
and axial mesons into the chiral Lagrangian. The Skyrme solution
will source these vector mesons through various cubic couplings such
as $g_{v^{(k)}\pi\pi}$, and in turn will backreact to these
classical excitations of vector fields. It just so happens that the
net result tends to shrink the size of the Skyrmion while preserving
the winding number. This tendency was also demonstrated some time
ago by incorporating $\rho$ meson in the chiral Lagrangian with
HLS$_1$ in the conventional field theory setting
\cite{vector-skyrmion}. What we found here is that in the strong
coupling limit with the entire tower of vector mesons included this
backreaction of the Skyrmion is rather extreme.

While we identified the instanton soliton as the carrier of baryonic quantum
numbers, we actually set up effective action for a subclass of baryons.
We restricted our attention to $N_F=2$ case, and considered dynamics of
the iso-doublet under $SU(N_F=2)$. These are of course the proton-neutron
pair. For these nucleons, we found a simple five-dimensional effective action where
all cubic and quartic interaction with mesons are encoded in two interaction
terms: a minimal coupling of the baryon current to $U(N_F=2)$ gauge field of the
form, $\bar{\cal B}A_{m}\gamma^{m}{\cal B}$, and a magnetic coupling to
$SU(N_F=2)$ gauge field strength of the form $\bar{\cal B}F_{mn}\gamma^{mn}{\cal B}$.
Considering that the gauge field includes the entire tower of vector,
axial-vector mesons, and pions this universal form of the interaction is simply
staggering.

Electromagnetic vertices are also extracted, more indirectly using
the AdS/CFT prescription relating source terms to the boundary
operators and bulk fields. The most prominent feature found here is
the vector dominance in a {\it generalized sense}. The
$conventional$ vector dominance refers to the assertion that photon
couples to hadrons only indirectly via mixing with the lowest lying
vector mesons, namely $\rho$ and $\omega$.
Here, instead, we showed explicitly that the photon field couples to
nucleons indirectly by mixing with the infinite tower of vector
mesons in the manner similar to the case of pions.

In contrast to the $conventional$ vector dominance which holds
poorly for the nucleons, photon has no direct contact coupling
with the nucleons (and pions). For small momentum transfer, where
our model is valid, in turns out that the first four vector
mesons, respectively in the iso-scalar and the iso-vector sectors,
dominate the form factors. While the vector dominance for pions is
relatively well-established even with the lowest vector mesons,
the vector dominance for nucleons has been more
controversial~\cite{bijker-iachello,iachello-jackson-lande,BRW-VD}.
A remarkable result of our findings is that while in the usual
Skyrme model, the direct photon coupling to the soliton is
mandatory~\cite{holzwarth}, in terms of the instanton, the direct
coupling can be transformed away and the full vector dominance,
albeit with the infinite tower, is recovered. One can think of
this as a ``derivation" of vector dominance model for the nucleon.

We devoted much of this article to exploring consequences of this
effective action, and made  effort to match with experimental data.
Qualitative predictions, such as large $N_c$ behaviors of chiral
couplings and ratios between various vector meson couplings, seem to
match with data fairly well, and general tendencies of subleading
corrections also concur with experiments. Upon some extra
assumptions on subleading corrections, motivated by CQM, quantities
like the axial coupling to pions and anomalous magnetic moment seem
to match rather well. We should however admit that so far our effort
to reach out to the experimental data is at best rudimentary. In
particular, the extraction of coupling constants are usually quite
model-dependent and we must fill the gap between the model and the
data by computing actual amplitudes, which would require going
beyond the large 't Hooft and large $N_c$ approximations. Thus a lot
more work is needed before our theory can confront the real data,
e.g., the precise JLab data on nucleon form factors etc.

Also, as a theoretical model, we have various improvements that are
still desired. In practice, the biggest hurdle in using our
effective action to its full potential ability lies with the
magnetic coupling $g_5(w)$. As we emphasized earlier, we have an
accurate number for its central value $g_5(0)$ only, owing to the
fact that the instanton solution exists only when centered at $w=0$.
The simple procedure we adopted in section 4 cannot be used to
extract $g_5(w\neq 0)$. The uncertainty due to this ignorance can be
minimized in the large $\lambda N_c$ limit whereby the baryon
wavefunction gets squeezed near the center $w=0$ along the fifth
direction, and the large $N_c$ limit of quantities like $g_A$ and
the anomalous magnetic moment are insensitive to this problem.
However, the extrapolation to small $\lambda N_c$ will be hampered
by this ignorance to various degrees, especially for quantities
whose dominant contribution arises from the magnetic term. We took a
simplifying assumption, $g_5(w)/e(w)^2= g_5(0)/e(0)^2$, for our
numerical estimates but this must be improved further.

Another immediate problem to address is the question of excited
baryons. The general approach we took should be certainly
applicable to more general baryons, such as $\Delta$, but the
precise form and the coupling constant of the magnetic term in
section 4 will be modified since the details of the latter
depended on the spin and iso-spin structure of the baryon field in
question. How the magnetic term will be modified for higher
iso-spin baryons remains unaddressed at the moment.

Finally, one would like to generalize the instanton picture to
hyperons. It is known that the conventional approach with the
Skyrme Lagrangian becomes inefficient when going from $U(2)$ to
$U(3)$. We should expect no better result with our model if we
tried to consider $U(3)$, especially since we do not know of natural
way to incorporate the strange quark mass.\footnote{Ref.~\cite{Casero:2007ae} studied
related issues in a more general D-branes/anti-D-branes setting
which allows bare masses of matter fermions from open string tachyon
field.} A possible approach to hyperons is the
Callan-Klebanov bound-state model~\cite{CK} where the kaons are
introduced as extra massive pseudo-scalars in doublets under
$U(N_F=2)$ and bound to the $SU(2)$ soliton. This approach was far
more successful than $U(3)$-based models, particularly if vector
mesons were included in the Lagrangian~\cite{rho,scoccolaetal}. It
would be interesting to work out the effective action for hyperons
as well as exotic baryon (e.g., pentaquark) structure with kaons
bound to the instanton.

\subsection*{Note Added}

The original version of this article used a non-canonical normalization
for vector mesons and axial vector mesons. In this latest version, we restored
the canonical normalization, which resulted in various multiplicative
factors in equations (2.26), (7.5), (7.6), (7.13), (7.17), and (7.20).
Also affected are some of the numerical entries for vector mesons in table 1 and
in table 2.

\subsection*{Acknowledgments}
 M.R acknowledges the
hospitality of the Department of Physics of Pusan National
University where this collaboration was initiated and is
particularly grateful to Chang-Hwan Lee for making the visit most
fruitful and pleasant. Three of us (D.K.H., H.U.Y., and P.Y.)
thank Shigeki Sugimoto and Koichi Yamawaki for helpful discussions
and the invitation to Nagoya University. H.U.Y. also acknowledges
interesting comments from Dongsu Bak and Soo-Jong Rey. P.Y. wishes
to thank Koji Hashimoto, Youngman Kim, Pyungwon Ko, Washington
Taylor, and especially Tae-Sun Park for useful conversations, and
is grateful to the String Theory Group at University of Tokyo,
where part of this work was done. We all thank Seonho Choi for his
JLab figures. This work was supported in part (D.K.H.) by KOSEF
Basic Research Program with the grant No. R01-2006-000-10912-0, by
the KRF Grants, (M.R.) KRF-2006-209-C00002, (H.U.Y.)
KRF-2005-070-C00030, and (P.Y.) by the KOSEF through the Quantum
Spacetime(CQUeST) Center of Sogang University with the grant
number R11-2005-021.

\end{document}